# Superionic diffusion through frustrated energy landscape

D. Di Stefano[1], A. Miglio[1], K. Robeyns[1], Y. Filinchuk[1], M. Lechartier[2], A. Senyshyn [3], H. Ishida[4], S. Spannenberger[5], D. Prutsch,[6] S. Lunghammer,[6] D. Rettenwander,[6] M. Wilkening,[6] B. Roling[5], Y. Kato[2]*, G. Hautier[1]*

**Affiliations:**

[1]Institute of Condensed Matter and Nanosciences (IMCN), Université catholique de Louvain, Chemin des étoiles 8, 1348 Louvain-la-Neuve, Belgium

[2]Battery AT, Advanced Technology 1, Toyota Motor Europe NV/SA, Hoge Wei 33A 1930 Zaventem, Belgium.

[3]Heinz Maier-Leibnitz Zentrum, Technische Universität München, 85748 Garching, Germany.

[4]Toray Research Center, Inc., 3-3-7 Sonoyama, Otsu, Shiga 520-8567, Japan.

[5]Department of Chemistry, Philipps-Universität Marburg, 35032 Marburg, Germany.

[6]Institute for Chemistry and Technology of Materials, Graz University of Technology, Stremayrgasse 9, A-8010 Graz, Austria

*Correspondence to:

geoffroy.hautier@uclouvain.be

yuki_katoh@mail.toyota.co.jp

**Abstract**:

Solid-state materials with high ionic conduction are necessary to many technologies including all-solid-state Li-ion batteries. Understanding how crystal structure dictates ionic diffusion is at the root of the development of fast ionic conductors. Here, we show that $LiTi_2(PS_4)_3$ exhibits a Li-ion diffusion coefficient about an order of magnitude higher than current state-of-the-art lithium superionic conductors. We rationalize this observation by the unusual crystal structure of $LiTi_2(PS_4)_3$ which offers no regular tetrahedral or octahedral sites for lithium to favorably occupy. This creates a smooth, frustrated energy landscape resembling more the energy landscapes present in liquids than in typical solids. This frustrated energy landscape leads to a high diffusion coefficient combining low activation energy with a high pre-factor.

The understanding and control of ionic conduction in solids is driving the development of many devices from sensors to fuel cells. In the field of Li-ion or Na-ion batteries, ionic transport in the electrode and electrolyte is often a bottleneck to higher rate and power. The development of superionic conductors for solid-state electrolytes has been a recent focus[1]. All-solid-state batteries would offer opportunities in terms of improved power, energy density and stability compared to Li-ion battery technologies based on organic liquid electrolytes[2, 3]. In addition, solid-state electrolytes could facilitate the development of next generation batteries, such as Li-S or Li-$O_2$ batteries[1]. However, very high ionic diffusion or conductivity in solids (i.e., comparable or close to liquids) at room temperature is an unusual phenomenon. Only a handful crystalline structure families have been reported as alkaline superionic conductors[4], such as beta-alumina[5], NASICON[6], garnet[7], Argyrodite[8], and $Li_{10}GeP_2S_{12}$ (LGPS)[9].

In the case of lithium conductors, the lithium ionic conductivity ($\sigma$) depends on the lithium conductivity (or charge) diffusion coefficient ($D_\sigma$) and the concentration of mobile lithium ($n_{Li}$) through $\sigma = n_{Li} D_\sigma \frac{q^2}{kT}$ where $q$ is the charge of the diffusing species, $k$ the Boltzmann constant and $T$ the temperature. The conductivity diffusion coefficient $D_\sigma$ is indicative of how easy lithium ions collectively flow when an electric field is applied. It is formally related to the tracer lithium diffusion ($D_{tr}$) coefficient by a factor called the Haven ratio $D_{tr} = D_\sigma H_r$. The Haven ratio for typical ionic conductor usually ranges from 0.1 to 1, while the tracer diffusion can vary by orders of magnitude between different materials. The tracer diffusion typically follows an Arrhenius law with $D_{tr} = D_0 e^{-\frac{E_a}{kT}}$. The activation energy ($E_a$) can be directly linked to the energy barriers necessary to jump between crystalline sites. Factors leading to superionic conductivity[10] include diffusion volumes and bond-valence path[11], phonon frequencies[12], face-sharing tetrahedral networks[13] or correlated cationic movements[14]. Since the beginning of the study of superionic conductors, the idea of developing solids with a frustrated energy landscape (i.e., for which no specific configuration for the mobile ions is favored) has been discussed, as it would lead to a diffusion behavior similar to that in liquids. In this regard, different types of frustration mechanisms have been recently put forward to explain high diffusion in garnets[15], argorydite[16], nanostructured $Ba_{1-x}Ca_xF_2$[17], multivalent cation electrodes[18] and borohydrides[19, 20].

Here, we show that $LiTi_2(PS_4)_3$ (LTPS) exhibits a very high Li-ion diffusion coefficient. The origin of the very fast diffusion is elucidated based on a detailed analysis of the frustrated energy landscape in this material.

We prepared samples of $LiTi_2(PS_4)_3$ (LTPS) using a solid-state reaction (see method in SI and **Fig. S1**). Powder X-ray diffraction (XRD) measurements confirm the purity of the previously identified LTPS phase (**Fig. S2**) [21–23]. We evaluated the lithium tracer diffusion coefficient in LTPS using $^7$Li pulsed field gradient (PFG) Nuclear Magnetic Resonance (NMR) at several temperatures (**Fig. S3 and Table S1**). **Fig. 1** compares the measured Li-diffusion coefficient versus temperature for LTPS (in red) to known solid (in blue) and liquid (in black) electrolytes[24–28]. A difference of more than one order of magnitude exists between liquid electrolytes and state-of-the-art solid-state electrolytes. This diffusion "gap" is partially filled by LTPS. LTPS exhibits a tracer diffusion coefficient, at room temperature, significantly higher (around $1.2 \times 10^{-11}$ m$^2$/s) than the best solid Li-conductors from the LGPS family. Its activation energy is around 246 meV and similar to LGPS (220 meV). The pre-factor ($D_0$) of $2.94 \times 10^{-7}$ m$^2$/s of LTPS is about 20 times higher than that of LGPS ($1.31 \times 10^{-8}$ m$^2$/s). We probed the conductivity of LTPS using impedance spectroscopy on pressed powder samples and found a lithium grain ionic conductivity of 6.1 mS/cm at 300K (**Fig. S4, S5**). An electronic conductivity of $8.2 \times 10^{-8}$ S/cm was estimated from DC polarization using an ion-blocking cell (**Fig. S6**). The activation energy of 277 meV extracted from impedance data is in agreement with the PFG-NMR data. Using the measured ionic conductivity and tracer diffusion, we evaluate an experimental Haven ratio around 0.24, on par with other solid-state electrolytes such as LGPS.

The grain conductivity of LTPS is already competitive with the best solid-state-electrolytes such as LGPS (12 mS/cm)[9]. The lower lithium concentration in LTPS compared to LGPS is compensated by its higher diffusion coefficient. The main limiting factor to the direct use of LTPS as a solid-state-electrolyte is the presence of $Ti^{4+}$ which is redox active in the voltage window of interest for Li-ion batteries. We note that computational results on the LTPS crystal structure where Ti was substituted by Zr (a non-redox active element) show very similar diffusion. This points to a diffusion mechanism strongly linked to the LTPS crystal structure (**Fig. S7**).

We performed extensive structural characterization to understand the crystal chemistry of LTPS using a combination of X-ray synchrotron and neutron powder diffraction, as well as XRD on a single crystal (**Table S2**). Kim et al reported a *P6cc* space group for LTPS [22, 23]. We find instead an orthorhombic *Ccc2* structure (**Fig. S2**), which is a superstructure of *P6cc*. These structures are very close each other. The orthorhombic *Ccc2* structure exhibits slightly lowered 3 fold point symmetry from Hexagonal *P6cc*. The structure presented in **Fig. 2 (a** and **b)** shows $TiS_6$ octahedra connected by edge sharing thiophosphate ($PS_4$) groups (**Fig. S8**). Through $^{31}P$ magic angle spinning (MAS) NMR we were able to resolve almost all magnetically distinct phosphorous sites in crystalline LTPS with chemical shifts smaller than 80 ppm (see **Fig. S9**). Narrow lines point to a well crystalline sample; the broad signals typically seen for thiophosphate glasses are absent. Therefore, we conclude that the crystalline structure of LTPS causes the high Li diffusivity probed by PFG NMR rather than any amorphous side phase whose amount seems to be negligible. Single crystal synchrotron X-ray diffraction experiments reveal Li ordering at 150 K (**Table S3**) and melting of the Li ions substructure at higher temperatures. In particular, Li positions cannot be localized on the same crystal at room temperature which is an indication of high lithium mobility. To gain further insight in the atomistic mechanisms controlling diffusion, we used *ab initio* molecular dynamics simulation (AIMD) within density functional theory (DFT). Experimentally unveiled space group *Ccc*2 for $LiTi_2(PS_4)_3$ was also independently obtained during a DFT optimization of *P*6*cc* initial model (see Structural Characterization in SI), and the relaxed structure show very closed structure to experimental one (**Fig. S10**). From the AIMD, the tracer diffusion can be extracted following the root mean square displacement of the lithium atoms with time (see SI). The AIMD provides an activation energy of 197 meV in fair agreement with experimental data (**Fig. S11, S12**). The Haven ratio can also be estimated using the AIMD data[14] and we find an Haven ratio of 0.3 in good agreement with experiment. The higher pre-factor present in LTPS compared to other superionic conductors such as LGPS is reproduced by our computational results (**Fig. S13**). While not constrained by symmetry, diffusion in LTPS is computed to be almost isotropic (**Fig. S12**). **Fig. 2C** shows an in-plane view of the LTPS crystal structure with Li-ion probability density obtained from the AIMD at two temperatures 1200K and 600K (more temperature are provided in **Fig. S14**). We can access the lithium diffusion paths using these probability density plots. It is noticeable that the large pore present in LTPS does not accommodate any lithium diffusion. Instead, lithium diffuses closely following the Ti-P-S framework forming "rings" around the Ti sites in the *a*-*b* plane. **Fig. S15** shows a tilted view of the structure indicating that those rings are also connected in the *c* direction leading to a 3D diffusion network. The lower temperature data (**Fig. 2c** 600K) shows the appearance of lithium "pockets" of high lithium probability. Three of these pockets constitute one "ring". The lithium ions jump between the pockets either within the ring (intra-ring jumps; blue arrows) or between rings (inter-ring jumps; red arrows). We observe from the molecular dynamic simulations that the fastest, higher rate, jumps are intra-ring and that the inter-ring jumps happen at a lower rate. The inter-rings jumps are the limiting steps for lithium macroscopic diffusion. Experimental complementary insight on this microscopic mechanism was also provided by $^{7}$Li-NMR relaxometry experiments (**Fig. S16**) which identify 2 jumps that we relate to the inter-ring and intra-ring processes. The inter-ring jump rate extracted from NMR relaxometry is in good agreement with our PFG-NMR diffusion coefficient.

We observe that the pockets of high lithium probability are not typical crystallographic sites (e.g, octahedral or tetrahedral) and are much larger "potato" shaped regions even at room temperature (see **Fig. 2d and Fig. S17** obtained from AIMD at 300K). This is unusual as alkali diffusion mechanisms in crystalline superionic conductors are typically described by jumps between connected crystallographic sites: tetrahedral-tetrahedral sharing faces for instance in the LGPS family or tetrahedral-octahedral-tetrahedral for garnets[13]. Observation of the Li-S coordination number for LTPS in these pockets shows that Li experience an average coordination between 3 and 4, far from the typical regular tetrahedral sites occupied for instance in LGPS (**Fig. S18**). This low coordination is rare as lithium sits in quite regular tetrahedral or octahedral sites in the vast majority of sulfides (**Fig. S19**). We link the unusual and energetically unfavorable coordination in LTPS and the occupation of large pockets to the absence of any regular tetrahedral (or octahedral) site in the Ti-P-S crystalline framework in which lithium could favorably sit. To test this hypothesis, we performed a local environment analysis of all (occupied by lithium and unoccupied) sites in LTPS to demonstrate the absence of undistorted tetrahedral sites favorable to lithium (Fig. S20). **Fig 3a** shows, in dashed black, a measure of the distortion of occupied Li tetrahedral environments in sulfide compounds from the stable materials listed in the Materials Project database (**Fig. S20, S21**)[29]. This analysis provides a range of the typical distortions that are acceptable for a tetrahedral site to favorably accept lithium in sulfides. The red curve in **Fig. 3a** shows the distribution of distortion for all tetrahedral sites in LTPS. The distribution of LTPS sites indicates that all available sites are highly distorted, providing no favorable sites for lithium to occupy. For comparison, the distribution of all available sites in LGPS is drawn in blue indicating a distribution range close to the one of sulfides present in the Materials Project. An analysis on octahedral sites leads to a similar conclusion (**Fig. S22**).

The crystalline Ti-P-S framework only provides very distorted tetrahedra for lithium to occupy. Since distorted tetrahedral sites are less energetically favorable for lithium (**Fig. S23** and **S24**), the lithium experiences in LTPS a frustrated energy landscape (i.e., for which no specific configuration for the mobile ions is favored). Additional evidence from sampling the LTPS framework with different configurations of lithium using DFT also indicates that the energy landscape felt by lithium is smooth and frustrated (**Fig. S25**). This frustration is mainly coming from the Ti-P-S framework and not from lithium-lithium interactions as indicated by the lack of dependence of the site energy distribution as well as the tracer diffusion on changes in lithium content (**Fig. S26**). This is different than the frustration described in other ionic conductors such as garnets[15].

In a simple model, two main interactions can be used to understand the energy landscape of lithium in an ionic conductor: Li-anion and Li-cation ($P^{5+}$, $Ti^{4+}$ in LTPS) interactions. The two components (Li-anion and Li-cation) are schematically described for a one-dimensional model of a traditional alkaline superionic conductor in **Fig. 3b**. The Li-anion interaction is modulated by the alternation of stable lithium sites separated by energy barriers due to the need for lithium to squeeze through a small polyhedral face or edge to reach the next stable site. Li-cation interactions are also present and tend to have a much longer modulation. In the resulting energy landscape, the energy barrier of lithium diffusion is mainly set by the Li-anion interaction. This energy landscape of typical superionic conductors is representative for instance of LGPS-like systems with Li-S landscape jump length around 2Å and a barrier set by lithium going through a face shared between two tetrahedra[13]. On the other hand, in the case of LTPS, the polyhedral sites are highly distorted leading to a smoother Li-anion component. To put it simply when there is no small polyhedral sites to jump between, there is no limiting small bottleneck (e.g., polyhedral faces) to go through. The resulting energy landscape is then mainly driven by the Li-cation interaction **(Fig. 3c)** and lithium occupies large pockets instead of small polyhedral sites (**Fig. 2d**). The position of these large pockets in the crystal structure is set by the Li-cation interaction. The Li-cation repulsion in LTPS is the lowest in the pockets regions of high lithium probability and the strongest at the diffusion bottleneck in the inter-ring region (**Fig. S27**). We note that the energy landscapes in traditional superionic conductors and LTPS differ significantly in their shape and by the importance of the Li-cation interaction (**Fig. 3b and 3c**).

Our result on LTPS shows that when the Li-anion energy landscape is smoothened, the energy barrier related to Li-cation interaction can, in the adequate material, lead to relatively small energy barriers (on par with the best current superionic conductors). The very large difference in the pre-factor of the tracer diffusion measured by PFG-NMR between LTPS and LGPS (around a factor 20) remains to be explained. We note that concerted lithium mechanism cannot be responsible for the higher pre-factor[14]. Indeed, the Haven ratio in LTPS is measured to be around 0.24 which is close to other superionic conductors and especially LGPS. This indicates that correlated movements are important in LTPS but not more than in other superionic conductors such as LGPS. According to transition state theory, the tracer diffusion coefficient for an interstitial mechanism can be expressed as $D_{tr} = \frac{1}{6}a^2 f \nu_0 z e^{\frac{\Delta S_{mig}}{k}} e^{\frac{-\Delta E}{kT}}$ where $a$ is the jump distance, $f$ a correlation factor, $\nu_0$ is the attempt frequency, $z$ is the number of neighbour sites to jump in, $\Delta S_{mig}$ is the entropy of migration, and $\Delta E$ is the energy barrier of migration. The pre-factor is directly influenced by the jump distance ($a$). To the contrary of traditional superionic conductors, the diffusion in LTPS is not limited to small jumps between polyhedra and the jump distance can be much longer (**Fig. 3b vs 3c, Fig. S28**). We estimate the jump length to be around 6Å in LTPS versus 2Å in LGPS accounting for a difference of 9 in pre-factor. It is unlikely that the correlation factor (f) could account for the remaining difference in pre-factor. The correlation factor cannot be larger than one and LGPS is reported to have a correlation factor close to one[24,25]. Additionally, an analysis of the AIMD simulation of LTPS identified only few inter-ring jumps returning backwards (see SI). NMR relaxometry provides a direct probe of the jump rate independently from the correlation factor and the jump distance. A comparison between the jump rate pre-factor $\Gamma_0 = \nu_0 e^{\frac{\Delta S_{mig}}{k}}$ in LGPS ($2.5 \times 10^{12}$ $s^{-1}$)[24, 25] and LTPS ($3.2 \times 10^{12}$ $s^{-1}$) extracted from the NMR data shows a higher pre-factor for LTPS (see $^7Li$ NMR spin-lattice relaxation measurements in SI) indicating that $\Gamma_0$ is also part of the higher tracer diffusion pre-factor in LTPS. As the stable sites in LTPS are large pockets (Fig. 2d), we expect softer vibrational modes than in LGPS as confirmed by a comparison of the vibrational density of states obtained from AIMD (see **Fig. S29**). The lower frequency vibrational lithium modes in LTPS lead to a lower attempt frequency (ν0) and entropically stabilized stable sites. Thus, the larger $\Gamma_0$ of LTPS can only be accounted for by a larger entropy of the transition state. **Fig. 3d** and **3e** compare the bond valence mismatch for lithium in the transition state for LGPS and LTPS. The region of lower bond valence mismatch indicates lower energy. LTPS shows a transition state significantly broader than in LGPS indicating that it is softer and thus entropically stabilized. This entropical stabilization of the transition state is inherent (as the larger jump length) to the frustrated energy landscape nature of LTPS since it comes from relaxing the constraint imposed to traditional superionic conductors of having to jump through narrow polyhedral edge or faces.

LTPS is not the only ionic conductor to present a high pre-factor for its diffusion coefficient. However, a high pre-factor is often correlated with a high energy barrier following the so-called Meyer-Neldel empirical rule. It is the combination of a high pre-factor and a low barrier that is rare in alkali superionic conductors and only seen in liquid electrolytes (see **Fig. S30**)[30]. Recently, Zeier et al. identified the Meyer-Neldel rule as a serious bottleneck to obtaining high performance superionic conductors.[31] The authors rationalized that the softening of the lattice which tends to lead to lower energy barriers is difficult to decouple from a lowering of the diffusion pre-factor through entropic stabilization of the stable site. Interestingly, LTPS keeps a high pre-factor despite low frequency soft modes through its long jump distance and entropic stabilization of the transition state. More generally, the high pre-factor, the unusual Li sites behavior (no occupation of a polyhedral sites) and the smoother energy landscape due to frustration are all characteristics of diffusion in liquids. In fact, the analysis of the lithium vibrational density of states during AIMD indicates that lithium mobility in LTPS resembles more a liquid than in LGPS (see **Fig. S29**). The reason for the high performance of some previously studied superionic conductors could potentially be rationalized by a similar frustrated energy landscape than in LTPS. In $RbAg_4I_5$, for instance, the mobile $Ag^+$ ions occupy large pockets instead of small polyhedral sites, and the $Ag^+$ ion conductivity is very high at room temperature due to a combination of high pre-factor and low energy barrier[32–34].

From its origin, the development of solid ionic conductors has been driven by the ambition to mimic in a solid the diffusion mechanism of a liquid. Most lithium superionic conductors (e.g., LGPS, garnets, …) show ionic conductivity mechanisms controlled by jumps between crystallographic sites (e.g., tetrahedral and octahedral). Here, we present a superionic conductor LTPS showing a deviation from these localized jumps as the lithium experiences a smooth, frustrated energy landscape. This frustrated energy landscape results from the unique crystal structure of LTPS offering no regular polyhedral sites for lithium to occupy. This unusual energy landscape influences the diffusion mechanism and leads to a very high lithium ion tracer diffusion due to a combination of a low energy barrier with a high pre-factor. This high pre-factor can be directly linked to long jump lengths and high entropy of the transition state. We believe that materials exhibiting a frustrated energy landscape, while rare, are likely to exist in other crystal structures than LTPS. Our work opens the possibility for searching for these exceptional crystalline frameworks through crystal structure analysis.

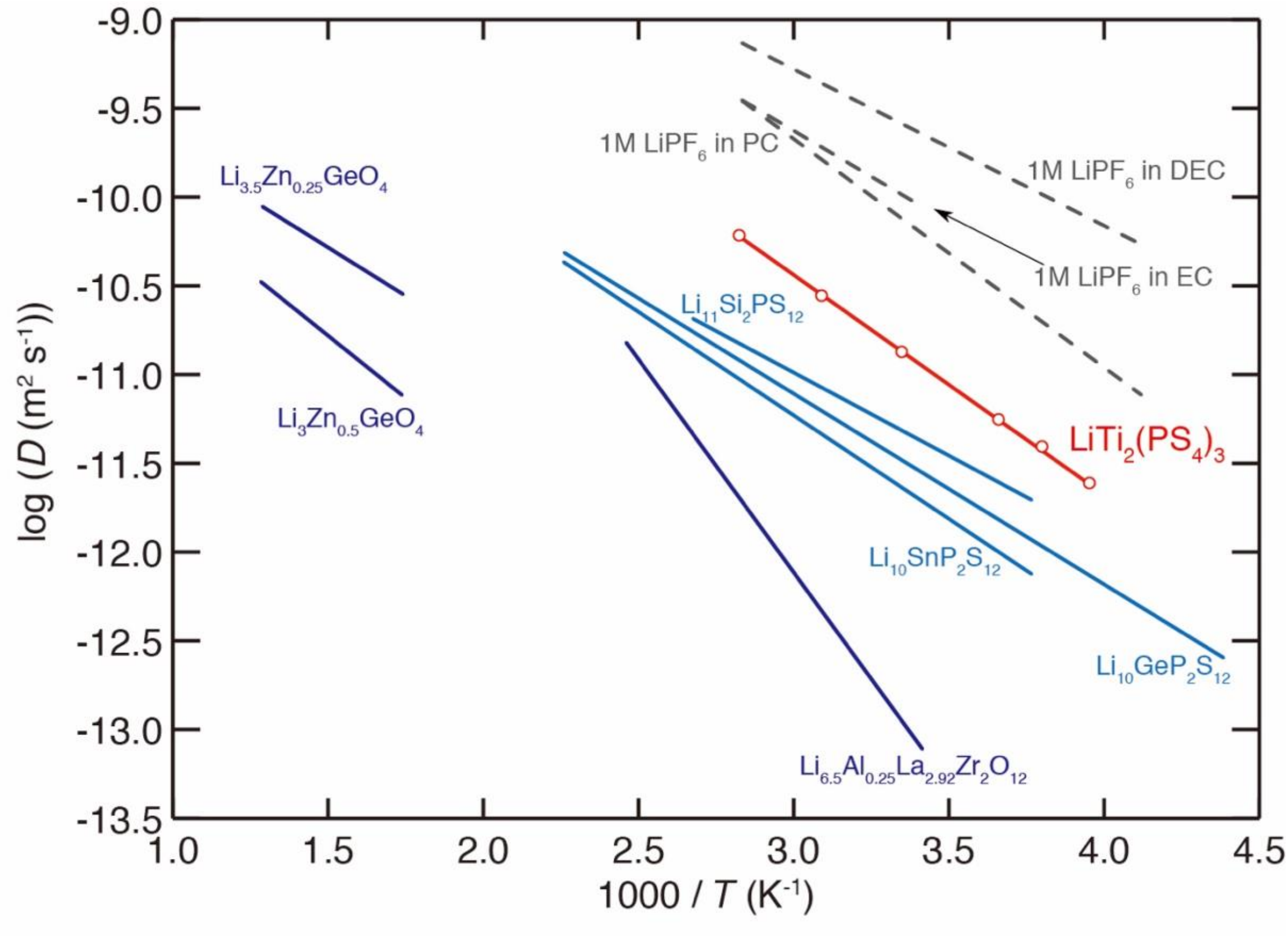

**Figure 1| PFG-NMR measured lithium diffusion coefficients versus temperature.** In blue colors for a series of solid electrolytes and in dashed black for liquid electrolytes. Our PFG-NMR measured results on LTPS are in reported in red.

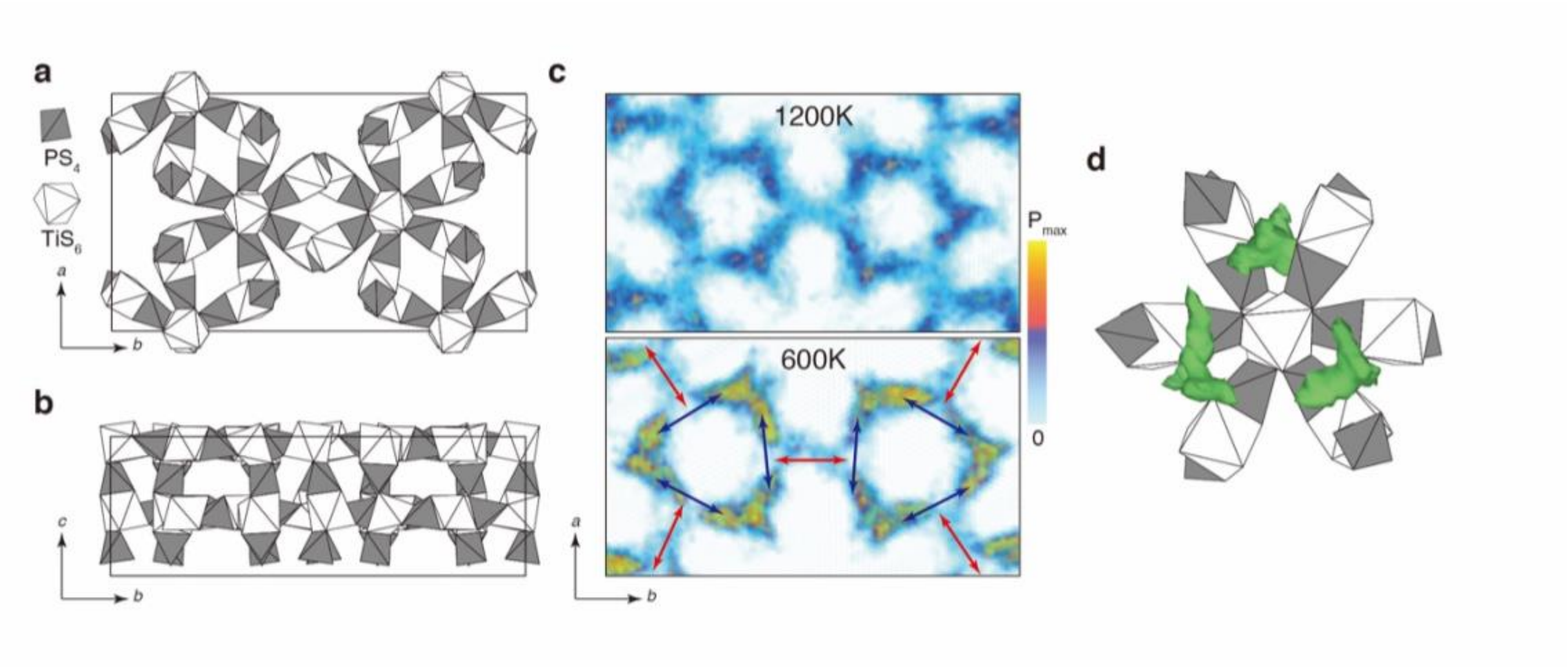

**Figure 2| LTPS crystal structure and lithium diffusion paths.** (**a** and **b**) Views of the LTPS crystal structure along the *c*, and *a* axis respectively. The lithium atoms are not shown.(**c**) Views along the *c* axis of the lithium probability density obtained during *ab initio* molecular dynamics simulation at 600K and 1200K. Blue arrows indicate the intra-ring jumps and red arrows the inter-ring jumps.(**d**) Regions of high lithium probability from AIMD simulations at 300K (obtained from 40 ps simulations)

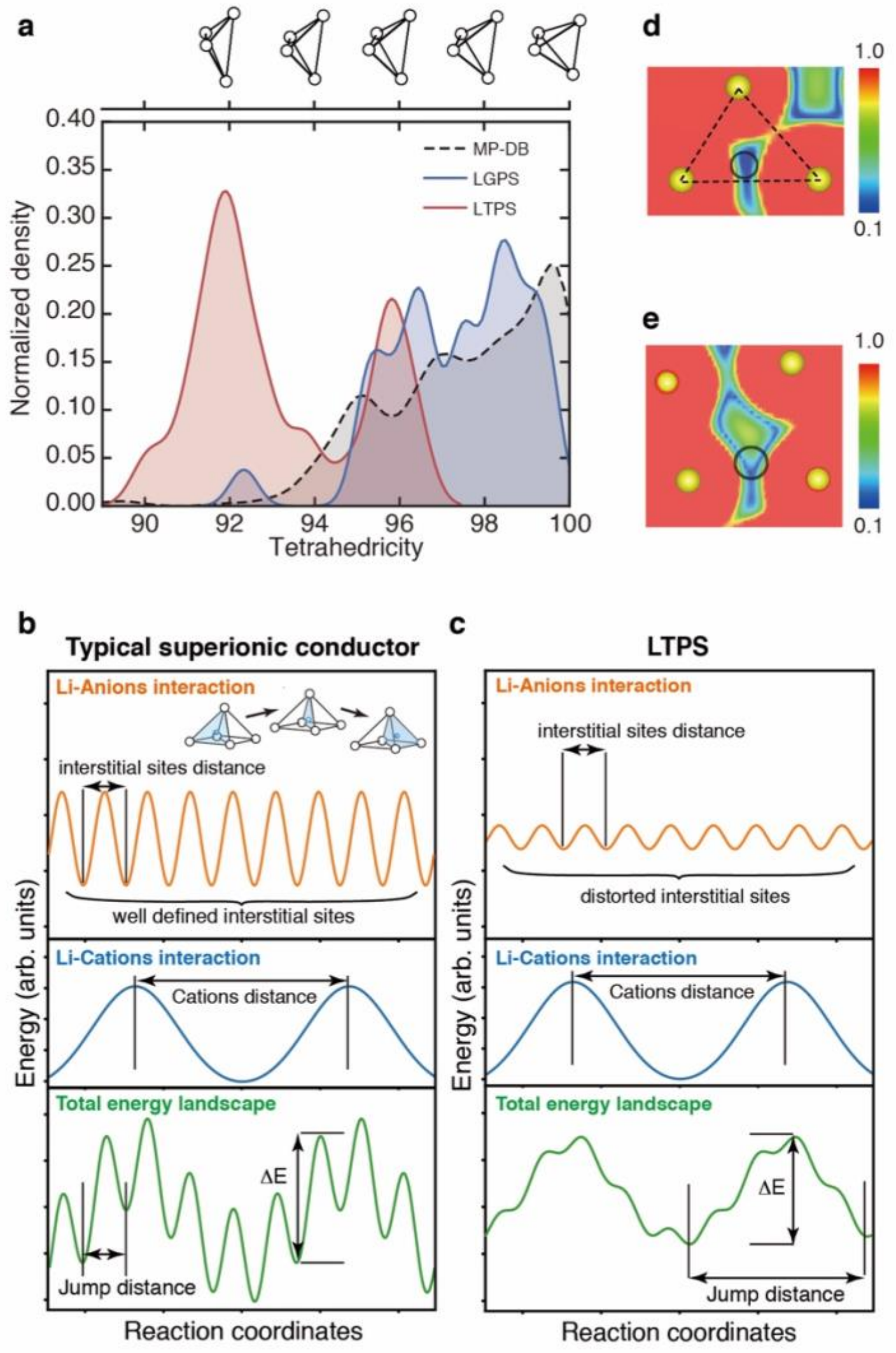

**Figure 3| Lithium sites distortion and diffusion mechanism in LTPS.** (**a**) In dashed black, tetrahedral sites distortion (tetrahedricity) distribution for the sites occupied by lithium in stable sulfides contained in the Materials Project database. In red and blue, tetrahedral sites distortion distribution for all (both occupied by lithium and unoccupied) sites in LTPS and LGPS, respectively. The tetrahedra on top of the figure indicate how tetrahedricity measures the distortion of the tetrahedron. (**b** and **c**) Lithium energy landscape for a model one-dimensional system including Li-anion and Li-cation interactions for a typical superionic conductor (e.g., LGPS) and for a frustrated energy landscape material (LTPS). (**d** and **e**) Differential bond valence mismatch for lithium (|1-Bond Valence (BV)|) at the transition state region for LGPS

(**d**) and LTPS (**e**). For reference we have reported the size of the LGPS transition state as dashed circle on the LTPS figure.

# Supplementary Information

## Superionic diffusion through frustrated energy landscape

D. Di Stefano, A. Miglio, K. Robeyns, Y. Filinchuk, M. Lechartier, A. Senyshyn , H. Ishida, S. Spannenberger, D. Prutsch, S. Lunghammer, D. Rettenwander, M. Wilkening, B. Roling, Y. Kato, G. Hautier

correspondence to: geoffroy.hautier@uclouvain.be and yuki_katoh@mail.toyota.co.jp

## Materials and methods

Synthesis:

The starting materials used for the synthesis of the $LiTi_2(PS_4)_3$ solid electrolyte were $Li_2S$ (99% purity, Sigma Aldrich), $P_2S_5$ (98% purity, Sigma Aldrich), and $TiS_2$ (99% purity, Sigma Aldrich). An enriched $^7Li_2S$ (Kojundo Chemical Lab.) was used as starting material for neutron diffraction measurement. All the reagents were weighed in the appropriate molar ratio and hand mixed with agate mortar. Then the mixture was treated by planetary ball milling (Fritsch, PL-7) for 1 hour at rotating speed of 370 rpm with $ZrO_2$ balls (ϕ10mm, 18 balls) using airtight 45 mL $ZrO_2$ pot. This milling procedure was repeated for 40 times with intermediate rest time of 15 min. The surface temperature of the $ZrO_2$ pot was measured by temperature label and it was below 60 °C during procedure. All procedures were conducted under an argon atmosphere ($H_2O$ < 0.1ppm and $O_2$ < 0.1ppm) inside a glove box (LABsatr, MBRAUN). The specimens were then pressed into pellets (ϕ8mm; applying 200MPa) and sealed in a ϕ9mm quartz tube together with slight excess of sulfur (99.998% purity, Sigma Aldrich) in order to make sulfur atmosphere inside of the quartz tube and to oxidize Ti completely. The polycrystalline material for PFG-NMR, impedance and powder XRD measurement was prepared on samples treated at 400 ºC for 8hours (heating rate 100ºC $h^{-1}$, natural cooling to R.T.). The heat treatment was conducted at the temperature between crystallization and decomposition (**Fig. S1**). We note that the previous synthesis route without ball milling[22, 23] was not able to offer the single phase of $LiTi_2(PS_4)_3$; mixture of $LiTi_2(PS_4)_3$ and $TiS_2$ was obtained. The single crystal for structural analysis was obtained as following: the quartz sealed precursor was heat treated at 750 ºC for 10 hours and slowly cooled to 400 ºC at a cooling rate of 3 ºC $h^{-1}$, then it was cooled to room temperature without temperature controlling. The single-crystals of $LiTi_2(PS_4)_3$, needle-like green metallic colored particles, were picked from the mixture of $TiS_2$ and decompositions.

Structural Characterization:

Preliminary X-ray diffraction studies were done using Laboratory X-ray powder diffractometers using $CuK_{\alpha 1}$ and $MoK_{\alpha}$ radiation. Phase-pure samples were identified. The Rietveld refinement using the hexagonal structure model published for $NaTi_2(PS_4)_3$ (ref 35) (the same *P*6*cc* structure was assigned to $LiTi_2(PS_4)_3$ (ref 23)) yielded substantially deformed $PS_4$ groups. This inconsistency compared to the expected tetrahedral geometry, as well as a significant peaks' overlap ($a/c$ is close to $\sqrt{3}$), urged us to use high resolution synchrotron powder diffraction.

Synchrotron radiation powder X-ray diffraction (SR-XPD) data were taken at the BL19B2 beamline at SPring-8 in Japan. The specimen was sealed in a quartz capillary of 0.5 mm diameter avoiding any contact with air. Diffraction data were collected in 0.01° steps from 2° to 78° in $2\theta$. The incident-beam wavelength was calibrated with $CeO_2$ and fixed at 0.5000 Å. The data were taken at 300 and 500 K. No phase transition was detected between these temperatures. Le Bail fit in the space group *P*6*cc* has shown the $a/c$ ratio of $\sqrt{2.9982}$, leading to practically complete overlap of Bragg peaks 002 and 300 and alike. To check for alternative unit cells, the indexing was attempted in Dicvol 2006.[36] High figures of merit were obtained for two different hexagonal cells, including one new and the one known for the *P*6*cc* structure model, as well as for two new orthorhombic cells. The analysis for systematic absences was hampered by the peaks' overlap, thus the space group determination was ambiguous, considering the need to verify for the supercells, even with the use of advanced tools such as TrueCell in the Chekcell program, version 2.[37] The Rietveld refinement in *P*6*cc* leads to a good fit but highly deformed $PS_4$ groups, similar to the results from the laboratory X-ray data.

Given the difficulties raised above, neutron powder diffraction (NPD) has been undertaken on $LiTi_2(PS_4)_3$ containing $^7Li$ isotope. The neutron diffraction data were collected at the high-resolution powder diffractometer SPODI at FRM-II reactor.[38] The sample was sealed in a 10 mm diameter vanadium cylinder under Ar using an indium ring. Diffraction data were collected in 0.05° steps from 1º to 150º in 2*θ*. Monochromatic neutrons were obtained using the 331 reflection from a composite Ge monochromator at a take-off angle of 155°. The wavelength was calibrated with NIST $LaB_6$, yielding 2.53612 Å. The exposure time was 16.5 hours, the temperature 300 K. In order to avoid possible preferred orientation, the data collection was performed under sample permanent rotation. Rietveld refinement of the *P*6*cc* structure to the NPD data was not satisfactory, especially at low angles. The apparent deformation of the $PS_4$ groups was in addition followed by an apparent disorder on the special Ti position. An improvement of the fit was not achieved by trying to complete the model with Li atoms located from difference Fourier maps.

The peaks overlap, poor fit and inconsistent molecular geometry questioned the structural model in the space group *P*6*cc*. However, indexing in alternative unit cells and space group determination were hardly possible given the extreme peaks' overlap even in the high-resolution powder diffraction patterns. Therefore, we attempted to grow single crystals of $LiTi_2(PS_4)_3$, see the procedure described above. Needle-like crystals with the average size of 0.2x0.06x0.06 mm were loaded in a 0.5 mm capillary under protective atmosphere inside an argon filled glove box (0.1 ppm of water and oxygen). The flame-sealed capillary was placed on the spindle of the MAR345 image-plate goniometer. The capillary was scanned to locate an isolated and suitable single crystal, which was properly aligned prior to the X-ray experiments. A total of 150 images with rotation steps of 1.5° were collected using MoKα radiation generated by a Rigaku UltraX 18S rotating anode equipped with Xenocs Fox 3D mirror at ambient temperature.

The single crystal data allowed to index the data in an orthorhombic cell with space group symmetry *Ccc*2. The data were integrated by CrysAlisPro version 1.171.37.35 (ref 39) and the implemented absorption correction was applied. The structure was solved by SHELXT[40] and refined by full matrix least-squares against $|F^2|$ using SHELXL-2014/7 (ref 41). **Table S1** combines the data collection and refinement statistics.

The crystal was found to be twinned and the TWINROTMAT procedure in Platon 2016 [42] was used to generate an HKLF5 formatted reflection file. Two twin matrices were selected along with racemic twinning. Refinement of the BASF factors indicated that the crystal is composed of three domains, the reported structure (24%) and its inverted structure (37%), and the inverted structure after rotation around the reciprocal (3 9 2) axis. The Li atoms could not be located in the crystal structure. The Squeeze procedure as implemented in Platon 2016 was used to estimate the volume and electron density in the voids and model this residual density in the subsequent refinement cycles. A probe of 1.2 Å radius was used during this procedure. The accessible volume is 1125 $Å^3$ per cell, of which 487 $Å^3$ is situated inside each of the two channels (per cell), extending along the *c*-axis. The number of "squeezed" electrons is 146 per unit cell and 68 per channel.

Interestingly, the unveiled space group *Ccc*2 for $LiTi_2(PS_4)_3$ was also independently obtained during a DFT optimization of the *P*6*cc* structure model done without symmetry constraints (i.e. in the space group *P*1) (**Fig. S10**). The Addsym procedure in Platon 2016 suggested the *Ccc*2 symmetry for the relaxed structure. The obtained structure is in the same 20x34x11 Å cell and clearly corresponds to the structure obtained by single crystal X-ray diffraction measurement.

Once the presumably correct structural model was obtained in the space group *Ccc*2, Rietveld refinements were done on the SR-XPD and NPD data using the program Fullprof 2016[43]. The atomic coordinates and the anisotropic displacement parameters were taken from the single-crystal structure and fixed in the refinements, varying only the scale factor, the cell parameters, the half-width fitted by the Caglioti equation, the shape modeled by pseudo-Voigt function, the low angle asymmetry, and zero shift. The background was described by linear interpolation between selected points.

The Squeeze procedure is not available for powder diffraction, prone to peaks' overlap, thus the electron and nuclear density in the channels had to be modeled in order to remove the significant intensity differences at low angles between the experimental data and the structure from the single crystal X-ray diffraction. Excellent fits can be achieved with two independent atomic scatterers (X1 and X2, see the **Table S4**): one in a general position and the other on the 2-fold axis. The refined atomic displacement factors are very high compared to the rest of the structure. They were refined iteratively with occupancies to convergence and then fixed. All three diffraction methods (single crystal XRD data not applying the Squeeze procedure, SR-XRD and NPD, **Table S4**) give consistent results.

Significant scattering power indicates it is not related to Li atoms but to inclusions, originating from the synthesis method. In particular, these may be sulfur atoms coming from the excess sulfur pelletized with the sample right before the annealing. The X atoms do not represent any particular element inside the pores, but merely allow to model the amount of electrons and nuclear density disordered inside the pores. Further study is needed to clarify the nature of these inclusions.

Highly accurate fits were obtained using the single crystal model with the powder SR-XRD and NPD data. For SR-XPD data, the final discrepancy factors, not corrected for background, are: $R_B$ = 8.8%, $R_F$ = 13.3%, $R_p$ = 4.2%, and $R_{wp}$ = 6.8%, $\chi^2$ = 13.2. The final refinement profile is shown in **Fig. S2** (top). The refined cell parameters are: $a$ = 20.0100(17), $b$ = 34.548(3), $c$ = 11.5352(7) Å, $V$ = 7974(1) Å$^3$. For NPD data, the final discrepancy factors, not corrected for background, are: $R_B$ = 11.6%, $R_F$ = 7.2%, $R_p$ = 2.6%, and $R_{wp}$ = 3.7%, $\chi^2$ = 6.1. The final refinement profile is shown in **Fig. S2** (bottom). The refined cell parameters are: $a$ = 20.018(2), $b$ = 34.563(3), $c$ = 11.5336(7) Å, $V$ = 7980(1) Å$^3$. In addition, the highly accurate fits of single crystal model with powder diffraction pattern indicate that ball milling pre-treatment has no effect on bulk structure of $LiTi_2(PS_4)_3$.

Fourier maps from all three methods do not allow to localize the Li atoms due probably to dynamic disorder. The occupancies of the Li sites found in the low energy DFT-relaxed structures also refine close to zero. Thus we conclude that the lithium positions cannot be experimentally determined at 300 K because of the extreme delocalization of lithium atoms.

Variable temperature single crystal synchrotron X-ray diffraction experiments:

Single crystals of the title compound were studied at the Swiss-Norwegian Beam Lines (SNBL) BM01A at the European Synchrotron Radiation Facility (ESRF) (Grenoble, France), using a PILATUS 2M hybrid pixel detector at a wavelength of 0.71490 Å. Crystals were loaded in glass capillaries under argon, some of them were transferred to dry mineral oil for less than a minute prior to data collection. A series of measurements was done on capillary-loaded crystals at various temperatures, starting from the room temperature and down to 100 K. Separate measurements on flash-frozen crystals in oil were done at 150 K. The temperature was controlled using an Oxford Cryostream 700+. All data show that at low temperatures (100 and 150 K) Li positions order, as they can be easily localized and refined as fully occupied sites. Instead, at higher temperatures as well as the room temperature, the Li substructure melts as the Li positions cannot be localized. This evolution is not the subject of this paper and will be published elsewhere. However, we present here the 150 K structure, characterized on a single crystal mounted in protective oil, showing the fully ordered Li positions. For this purpose, 720 images were collected with 0.5° steps at a sample-to-detector distance of 141 mm. The data were converted and integrated using the SNBL toolbox 2015 software and the CrysAlisPro software, respectively[44]. The crystals were twinned similarly to those measured in the Lab, the structure model was refined as described above. **Table S3** contains the summary of the experiment and the refinement.

$^{31}$P NMR characterization

$^{31}$P magic angle spinning (MAS) NMR measurements were carried out on a Bruker Avance 500 MHz spectrometer. We used a commercial 2.5-mm MAS NMR double-resonance probe designed by Bruker. The $ZrO_2$ rotor was packed with the powder sample in Ar atmosphere ($O_2$ < 0.1 ppm, $H_2O$ < 0.1 ppm) to protect the sample from reaction with any moisture. The magnetic field of 11.7 T corresponds to a Larmor frequency of 202.4 MHz. We used single pulse excitation, the π/2 pulse length was 1.33 μs. 16 scans were accumulated to obtain the spectrum shown here, which was recorded at a spinning frequency of 25 kHz with ambient bearing gas. The recycle delay between each scan was 300 s. Chemical shift were referenced to crystalline $LiPF_6$ serving as a second reference (−150 ppm, primary reference 85% $H_3PO_4$).

$^{7}$Li-PFG-NMR characterization

$^{7}$Li pulse field gradient (PFG) NMR measurement was performed at 155.6 MHz using Avance III HD spectrometer (Bruker Biospin). A Diff60 diffusion probe was used for the measurement. The lithium tracer diffusion coefficient $D_{tr}$ was measured in diffusion time $\Delta$ = 6 ms to 500 ms at the temperature range of 253K to 353K (−20°C to 80°C) by varying the strength of pulse field gradient $g$ and its duration $\delta$ between 0 to 25 $Tm^{-1}$ and 1 ms to 2.5 ms, respectively.
Echo-attenuation curves obtained at different diffusion time $\Delta$ at 25 ºC are shown in **Fig. S3a**. Up to $\Delta$=25ms, the logarithm of Echo-attenuation curves were parameter ($g$, $\delta$, $\Delta$) independent forming a straight line (coefficient of determination $R^2$>0.995), indicating that diffusion is understood by simple tracer diffusion in a homogeneous system. We assign the measurements with $\Delta$<25ms with intra-crystalline diffusion. In this situation, tracer-diffusion coefficient ($D_{tr}$) can be directly estimated by using the Stejskal-Tanner equation[45–47]:

$$\ln\left(\frac{I}{I_0}\right) = -\gamma^2 D_{Tr}\delta^2\left(\Delta - \frac{1}{3}\delta\right)g^2$$

The lithium tracer diffusion coefficient is then estimated at different Δ with high fitting accuracy of standard deviation $S.D._{\mathrm{fit}} < 10^{-13}$ $m^2s^{-1}$. By using mean value $\overline{D}_{tr} = \frac{D_{tr_1}+\cdots+D_{tr_N}}{N}$ and standard error $S.E._{\Delta} = \frac{S.D._{\Delta}}{\sqrt{N}}$, ($S.D._{\Delta}$ is the standard deviation for different Δ), the true value of tracer diffusion coefficient is statistically estimated to lie in the range of $\overline{D}_{tr} - 3SE_{\Delta} < D_{tr} < \overline{D}_{tr} + 3SE_{\Delta}$ with 99.7% probability. The standard error in our measurement is within 1.2%. We obtained $D_{tr} = (1.30 \pm 0.036)\times10^{-11}$ $m^2s^{-1}$ at 25ºC.
Plotting the tracer diffusion in function of the root-mean square displacement as in **Fig. S3b** indicates a change around 1.0 μm which corresponds with the grain size estimated by diffraction. This agrees with our assumption that the data below a Δ of 25 ms relates to intra-grain diffusion.
The echo-attenuation curves relevant to intra-grain diffusion and at different temperatures from −20ºC to 80 ºC are shown in **Fig. S3c**. The extracted intra-grain lithium tracer diffusion coefficients are provided in **Table S1**.

Impedance spectroscopy

In order to reduce the grain boundary resistance, the pellet was densified using the following procedure. The ball-milled starting material was hot pressed at a temperature of 200 ºC for 10 min by applying 270 MPa pressure under Ar atmosphere. Then the obtained pellet (6 mm diameter) was sealed into quartz tube with slight excess of sulfur in order to obtain an oxidative atmosphere and heat treated at 400 ºC for 8h. To avoid the direct contact of additional sulfur and LTPS, sulfur was placed on the other side of LTPS in bended quartz tube. The excess sulfur can be condensed during cooling by the natural temperature gradient of furnace. In contrast, annealing under vacuum condition did not result in a well densified pellet; slightly decreased relative density indicated the loss of sulfur. The effect of sintering condition on impedance is shown in **Fig.S4**.

Ion-blocking Au electrodes were sputtered on both side of the pellet, and the pellet was mounted into an air-tight cell. The ac impedance of $LiTi_2(PS_4)_3$ was measured in a frequency range from 1 MHz to 0.1 Hz and at temperatures between −130 ºC to 27 ºC using a Novocontrol Alpha-AK impedance analyser equipped with a Quatro cryosystem. The applied ac voltage was 10 mV.

**Fig. S5a** shows the frequency dependence of the real part of the conductivity, $\sigma'$, measured at various temperatures. We found three conductivity plateaus and three dispersion steps, indicating three different types of conduction processes. The corresponding impedance data were analyzed by fitting of Nyquist plot at various temperatures (**Fig. S5b**). Inset of **Fig. S5b** shows the equivalent circuit used for fitting. Each of the three impedance semicircles is represented by a parallel combination of a resistor $R_i$ and a constant phase element $CPE_i$ (i=1,2,3 for semicircle i). From the fitting results, the capacitances of semicircles were determined as: $C_1 \approx 15$ pF cm$^{-2}$, $C_2 \approx 500$ pF cm$^{-2}$, and $C_3 \approx 30$ nF cm$^{-2}$. These values are indicative of (1) grain ion transport, (2) Maxwell-Wagner (MW) polarization, and (3) grain-boundary ion transport. The MW polarization is most likely due to a secondary phase generated at the grain boundaries during the densifying process (hot-press and sintering). Since no clear evidence for a secondary crystalline phase was found in XRD and no evidence for a thio-phosphate glassy phase was found in $^{31}$P MAS NMR, we expect the secondary phase to be either amorphous titanium sulphide or sulfur-defective LTPS.

The grain conductivity $\sigma_g$ and the total conductivity $\sigma_{total}$ were calculated as follows: $\sigma_g = (1 / R_1) * (L / A) * (1 / x)$ and $\sigma_{total} = (1 / (R_1+R_2+R_3)) * (L / A)$. Here, $L$, $A$ and x denote the thickness of the pellet, the area of the pellet, and the volume fraction of conductive phase, respectively. Since the MW polarization is highly pronounced, we expect that there is considerable amount of less conductive phase included in the pellet. However, since the actual volume fraction of the conductive $LiTi_2(PS_4)_3$ is not clear, we use x = 1 for the conductivity estimation. **Fig. S5c** shows the temperature dependence of the conductivity for $\sigma_g$ and $\sigma_{total}$. Both of the grain and total conductivity of $LiTi_2(PS_4)_3$ pellet show an Arrhenius behavior. The grain conductivity $\sigma_g$ exhibits an activation energy, $E_a$ = 277 meV. Extrapolation of the Arrhenius fit to 300 K results in a value of $\sigma_g$=6.1 mS cm$^{-1}$. The electronic partial conductivity was measured by DC polarization method using ion blocking electrode system; the LTPS pellet was sandwiched by stainless steel current collector and 100mV was applied during transient current measurement (**Fig. S6**). The electron conductivity was estimated to be $8.2\times10^{-8}$ Scm$^{-1}$ at 27ºC. The transference number of electron $t_e$ was estimated as $t_e = \sigma_e / (\sigma_{ion} + \sigma_e) = 1.3\times10^{-5}$. Using the lithium tracer diffusion obtained from PFG-NMR and the grain conductivity obtained from impedance, we calculate the experimental Haven ratio $H_R$ as follows:

$$H_R = \frac{D_{tr}}{D_\sigma} = \frac{D_{tr}}{\frac{k_B \cdot T}{N_{Li} q^2}\sigma_g}$$

where $k_B$, $q$, and $N_{Li}$ are Boltzmann's constant, the elementary charge, and the number density of mobile lithium in the $LiTi_2(PS_4)_3$ structure, respectively. $N_{Li} = 2.0\times10^{21}$ cm$^{-3}$ was determined using the crystallographic data: lattice parameters obtained at 300K and mobile lithium from the refined structure at 150K. We note that the lithium ions were not observable by diffraction at room temperature, but low temperature measurements (<150K) make the localization of lithium ions possible leading to the determination of the lithium content. The experimental Haven ratio in LTPS is measured to be 0.24 at room temperature.

Due the resistive nature of the grain boundaries of LTPS, the total ionic conductivity is $\sigma_{total}$ = 0.25 mS/cm at 300K. Such resistive grain boundaries are often found in superionic conductors, such as garnet- and NASICON-based systems, and can be reduced by optimizing the sintering conditions.[48, 49]

<u>Density Functional Theory computations and molecular dynamics</u>

All ab initio computations in this study were performed using density functional theory within the Projector Augmented Wave (PAW)[50] approach as implemented in the VASP code version 5.4.1.(ref 51)

PAW datasets with the Perdew Burke Ernzerhof (PBE) [52] functional were used to calculate the exchange-correlation terms of Li ($1s^2$ $2s^1$ PAW_PBE Li_sv 23Jan2001), Ti ($3p^6$ $4s^2$ $3d^2$ PAW_PBE Ti_pv 07Sep2000), Zr ($4p^6$ $5s^2$ $4d^2$ PAW_PBE Zr_sv 04Jan2005), P ($3s^2$ $3p^3$ PAW_PBE P 17Jan2003) and S ($3s^2$ $3p^4$ PAW_PBE S 17Jan2003) valence electrons.

The molecular dynamics simulations were performed using a Verlet algorithm to integrate Newton's equations of motion, at a time step of 2 fs for a total simulation time of at least 200 ps on an NVT ensemble where we fix the number of atoms, the volume and the temperature[53–55]. We start all runs from a temperature of 100 K, rise it to the target temperature over a time period of 2 ps and then equilibrate the system at the target temperature for a time period of 10 ps before to start to sample diffusion data. A unit cell containing 144 atoms (8 formula units) and a Γ-point only k-point sampling were chosen. No melting or breaking of P-S or Ti-S bonds were observed during the simulations. Simulations at a lower time step (1fs) for 40 ps at 600K, 800K and 1200K did not show any significant difference in computed tracer diffusion. All analyses were performed using pymatgen 4.2.2 and the pymatgen-diffusion package.[56] For LGPS, we used a unit cell containing 50 atoms, Γ-point only k-point sampling and performed NVT AIMD simulations for at least 200 ps with a time step of 2 fs.

The AIMD simulations were performed in a temperature range of 600 to 1200 K. The tracer diffusion coefficient ($D_{tr}$) of each ion is then given by the slope of the mean squared displacement (MSD):

$$D_{tr} = \lim_{t \to \infty} \frac{\frac{1}{N}\sum_i^N < [\boldsymbol{r}_i(\mathrm{t}) - \boldsymbol{r}_i(0)]^2 >}{6t}$$

Where t is the simulation time, 6 comes from the 3D nature of the system and r is the position of a lithium atom. The average (< >) is performed on time and N is the number of lithium in the cell. The tracer diffusion coefficient is obtained from the slope of the (smoothed) MSD with time at each temperature (**Fig. S11**). The activation energy $E_a$ was obtained from an Arrhenius plot (**Fig. S12**).

The conductivity (or charge diffusion) ($D_\sigma$) is obtained from the AIMD as well.[14]

$$D_\sigma = \lim_{t \to \infty} \frac{< \left[\frac{1}{N}\sum_i^N \boldsymbol{r}_i(\mathrm{t}) - \frac{1}{N}\sum_i^N \boldsymbol{r}_i(0)\right]^2 >}{6t}$$

From the ratio between tracer diffusion and charge diffusion, we obtain the Haven ratio. The computed Haven ratio is around 0.3 in LTPS.
The number of jumps between rings was also analyzed in the AIMD at 1000K. On 65 inter-ring jumps detected, we found 8 that were followed by a jump backward (12%). This is a qualitative indication of an individual correlation factor (f) that deviates only slightly from one.
For atomic relaxations, the cell parameters and atomic positions were optimized with a plane wave cutoff energy of 520 eV and the residual forces were ≤ 0.01 eV $Å^{-1}$.

$^7$Li NMR spin-lattice relaxation measurements

$^7$Li (spin-3/2) spin-lattice relaxation rates in the laboratory frame were recorded with a Bruker 300 Avance spectrometer in combination with a static broadband probe (Bruker) which operated at a Larmor frequency of $\omega_0/2\pi$ = 116 MHz. We took advantage of the well-known saturation recovery pulse sequence to record spin-lattice relaxation rates $R_1$, see refs. [57, 58] for details. A train of 10 90° pulses (2.5 μs in length) destroyed any longitudinal magnetization in thermal equilibrium; its recovery as a function of waiting time was then recorded until full recovery had been achieved. The magnetization transients were parameterized by stretched exponentials to extract diffusion-induced spin-lattice relaxation rates $R_1$. The stretching exponents ranged from 1 to 0.8, thus showing only slight deviations from simple exponential recovery. Static $^7$Li NMR line shapes were recorded after excitation the spin ensemble with a single 90° pulse; the recycle delay was at least 5 × $1/R_1$ to ensure quantitative lines.
In **Fig. S16a** the diffusion-induced spin-lattice relaxation rates $R_1$ are shown in an Arrhenius plot to illustrate their behaviour with temperature. Usually the rate passes through a peak; at the temperature where the maximum shows up ($T_{max}$) the mean Li jump rate can be extracted in a rather model-independent approach [59–61]. At $T_{max}$ the Li jump rate $1/\tau$ is expected in the order of the Larmor frequency $\omega_0$ used to record the rates: $\omega_0\tau \approx 1$. In the present case the rates could be best parameterized with two diffusion-induced rate peaks. The solid line shows a sum of two BPP-type Lorentzian-shaped peaks (1 and 2); the dashed lines indicate the two peaks separately. The rate peaks show up at $T_{max}$ = 340 K and $T_{max}$ = 214 K, respectively. At these temperatures the jump rate associated to the jump process behind is in the order of $\omega_0$, i.e., the average residence time of a Li is in the ns regime. Using the information from the AIMD, we can assign peak (2) to successful jumps between the rings (inter-ring jumps). The faster process (1), seen as a shoulder of the main peak at lower temperatures, could be assigned to the intra-ring jumps (jumps between pockets). To our knowledge, peak (1) represents one of the fastest exchange processes seen by NMR relaxometry so far. The solid line shows the sum of two BPP fits using Lorentzian-shaped spectral density functions, (see, e.g. ref 58) as a guide to the eye. The dashed lines illustrate the individual rate peaks of the two diffusion processes.
The extremely fast Li exchange process in LTPS is further confirmed by $^7$Li NMR line shape measurements (**Fig. S16b**). In general, the central transition of the Li NMR line shape is dipolarly broadened at low *T*. Only sufficiently fast Li diffusion processes are able to average dipole-dipole interactions causing pronounced motionally induced line narrowing with increasing temperature. Here, at approximately 240 K, the NMR line has already reached the extreme narrowing regime indicating rapid ionic motion. At 293 K the line width (full width at half maximum, fwhm) is given by only 373 Hz. The residual line width originates from small field inhomogeneities of the experimental setup. In agreement with such fast exchange processes a distinct (first order) quadrupolar powder pattern emerges at elevated *T*. The pattern is, if we simply assume an axially oriented mean electric field gradient, characterized by a coupling constant of 14.6 kHz. Another set of quadrupole singularities is seen close to the central transition. It indicates that the Li ions occupy several electrically inequivalent sites in LTPS that take part in overall diffusion. At temperatures lower than 160 K we expect the line to broaden and to reach the so-called rigid lattice regime; in that regime the line width is usually given by several kHz.

The jump rate measured by NMR relaxometry is

$$\Gamma = \Gamma_0 \, e^{\frac{-\Delta E}{kT}} = \nu_0 e^{\frac{\Delta S_{mig}}{k}} e^{\frac{-\Delta E}{kT}}$$

At 340K, the peak (2) (inter-ring) jump rate is $7\text{x}10^8$ $s^{-1}$. Using the ΔE from PFG-NMR (246 meV), we can deduce $\Gamma_0 = 3.2\text{x}10^{12}$ $s^{-1}$ for LTPS. Similarly, using the data from Kuhn et al. (refs), we find a $\Gamma_0$ =$2.5\text{x}10^{12}$ for LGPS. Combining a jump distance of 6A with a correlation factor of 1, we can estimate a tracer diffusion of $1.4\text{x}10^{-11}$ $m^2/s$ (at 300K) from relaxometry. This is comparable to the PFG-NMR measured value of $1.2\text{x}10^{-11}$ $m^2/s$.

Local environment analysis:

The distortion of a local environment (LE) with respect to a model structure, such as a perfect tetrahedron or a perfect octahedron, can be measured (independently from the size and orientation) using the so-called continuous symmetry measure (csm) [62] as implemented in pymatgen 4.2.2 (chemenv package).[56] For the sake of clarity in this work we define the *tetrahedricity* (*octahedricity)* of a LE as 100-$csm_{T(O)}$ where $csm_{T(O)}$ is the csm calculated with respect to a perfect tetrahedron (octahedron). Since the csm by definition takes values from 0, for perfect match, to 100, for completely distorted shape, a tetrahedricity of 100 corresponds to a perfect tetrahedron while a tetrahedricity of 0 corresponds to a completely distorted one. We note that a tetrahedral site in a sulfur bcc anionic framework will have a tetrahedricity of about 98%.

To obtain the distribution of the distortion of occupied Li tetrahedral LEs in sulfide compounds on the entire Materials Project database, blue curve in **Fig 3a**, we calculated the tetrahedricity for each tetrahedral Li LE in each stable sulfide present in the Materials Project database on March 2017. In an analogous way, we have calculated the distribution of distortion for octahedral LEs shown in **Fig. S22**, in dashed black.

The red and blue curves in **Fig. 3a** (for tetrahedral LEs) and **Fig. S22** (for octahedral LEs) represent the distortion distribution on all available sites, occupied *and* unoccupied in LTPS and LGPS. For the occupied sites, it is enough to calculate the tetrahedricity/octahedricity for each Li LE as we did for all the stable Li-containing sulfides in the MP database. In order to identify the unoccupied sites is LTPS and LGPS and calculate their csm, we have developed the algorithm illustrated schematically in **Fig. S20**. Note that in the last step of the algorithm we removed the sites that are out of the typical volume range for $LiS_4$ tetrahedron or $LiS_6$ octahedron. This is necessary to avoid to consider sites that are too small, or too large, to accommodate favorably a Li atom. The acceptable volume range has been set for tetrahedral LEs to be from 7 to 8.5 Å$^3$ by analyzing the actual volume of each Li occupied tetrahedron in any Li containing sulfide in the MP database. In an analogous way, we set the volume range for octahedral LEs from 20 to 33 Å$^3$. The obtained volume distributions are shown in **Fig. S21 (a and b)** for tetrahedral and octahedral LEs, respectively.

Relation between site distortion and frustration

In order to support the causal connection between the distortion of a tetrahedron the energetics of the lithium occupation of this tetrahedral site, we have performed several model calculations which show the correlation between site tetrahedricity and energetics.

First, we have considered the formation energy of an isolated $LiS_4$ tetrahedron. The formation energy is defined as:

$$E_{form}^{d} = E\left(LiS_4^d\right) - E\left(S_4^d\right) - E_{ref}$$

Where $E\left(LiS_4^d\right)$ and $E\left(S_4^d\right)$ are the energy of the distorted sulfur tetrahedron with and without lithium, respectively. As reference value, we have chosen $E_{ref} = E(LiS_4^0) - E(S_4^0)$ with the 0 superscript indicating undistorted (tetrahedricity=100%) tetrahedron. The computations have been performed using the same parameters (e.g., pseudopotentials) as previously described in the subsection "Density Functional Theory computations and molecular dynamics", a cubic supercell with edge of 15 Å, a 8x8x8 k-point mesh, and a charge compensation to keep sulfur in a -2 and Li in a +1 formal oxidation state. In **Fig S23** we have plotted the formation energy vs tetrahedricity for this isolated $LiS_4$ tetrahedron which are strained uniaxially and by shear.
A second set of calculation was performed considering a model system of 54 S atoms arranged in a bcc framework under various uniaxial and biaxial strain distortions. For each obtained structure, we have introduced a Li atom in a tetrahedral site and let the ionic positions relax. The simulations have been performed using the same parameters (e.g., pseudopotentials) as previously described in the subsection "Density Functional Theory computations and molecular dynamics", a 2x2x2 k-point mesh, and a charge compensation to keep sulfur in a -2 and Li in a +1 formal oxidation state. **Fig S24** shows the tetrahedricity vs energy during the relaxation for a series of distortions of the bcc framework.

Calculation of the electrostatic potential

The electrostatic potential has been computed from Ewald sum as implemented in the pymatgen 4.2.2 code.[56, 63] Since we were interested in the Li-cations interaction, we have considered the positions of all cations other than Li and assigned to each of them their formal oxidation states ($P^{5+}$, $Ti^{4+}$). No Li-Li interaction is included here.

Velocity autocorrelation functions and Li vibrational density of states

The velocity autocorrelation functions have been obtained as:

$$Z(t) = \frac{1}{N}\sum_{i}^{N} < \boldsymbol{v}_i(0) \cdot \boldsymbol{v}_i(t) >$$

where $N$ is the number of Li in the structure, $<..>$ denotes the a time average, and the $\boldsymbol{v}_i(t)$ are the velocity of the Li atoms along the trajectories obtained from the AIMD runs. The vibrational density of state (vDOS) has been obtained as the direct Fourier transform of the velocity autocorrelation function. The results have been extracted from 50 ps AIMD simulations at 300K and 600K.

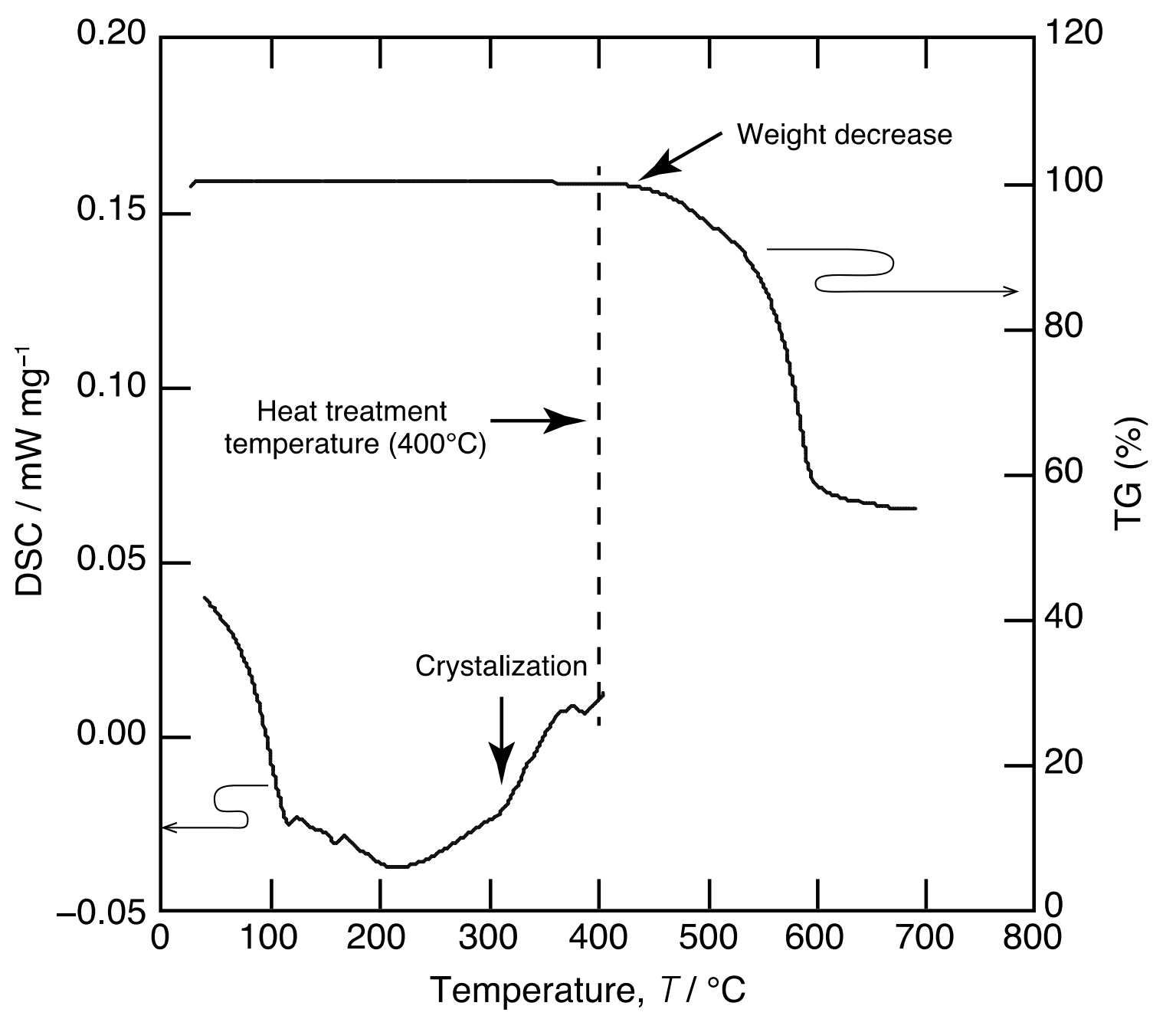

**Fig. S1. DSC and TG measurement of ball milled precursor of $LiTi_2(PS_4)_3$.**

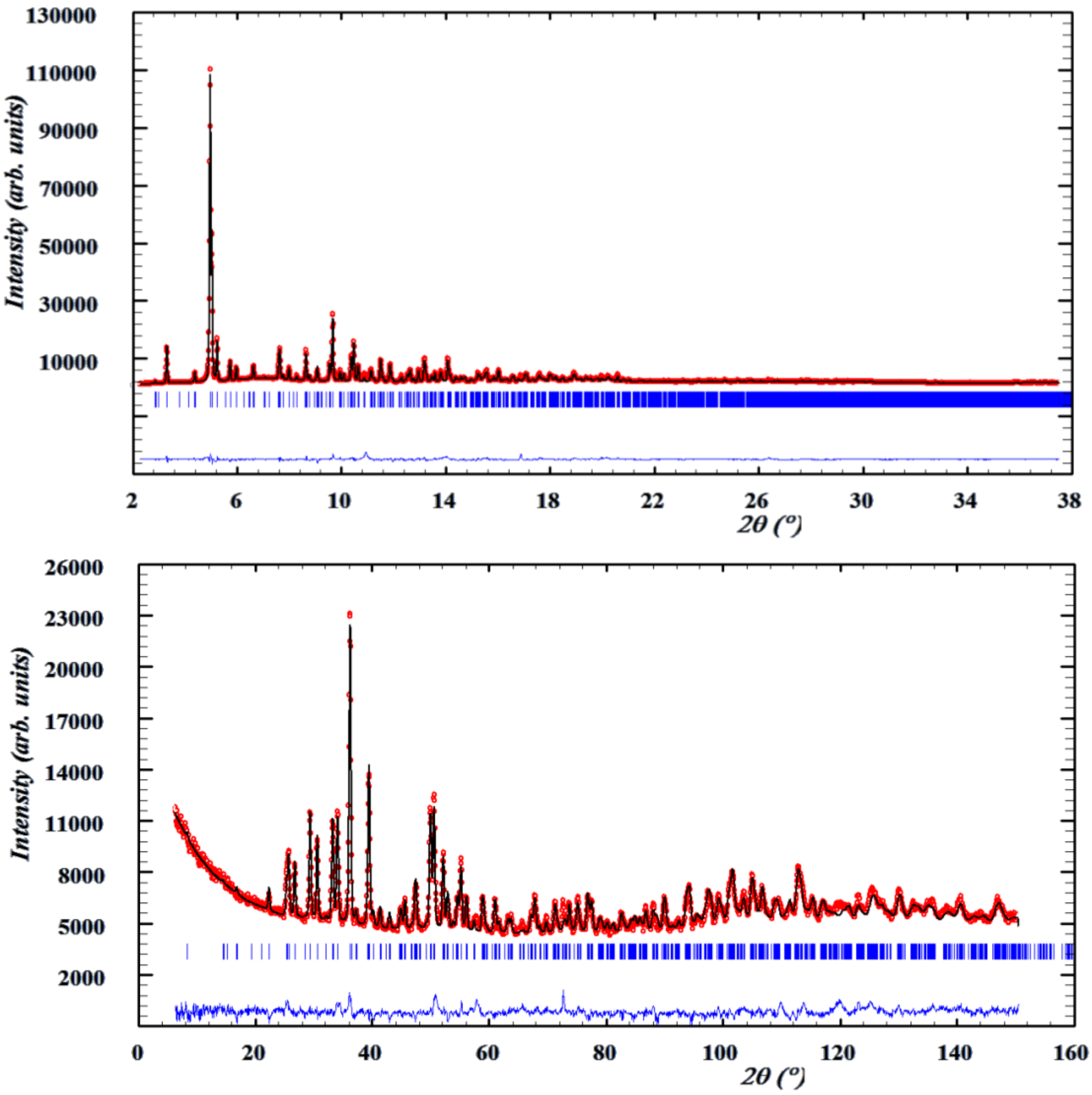

**Fig. S2| Rietveld refinement profile for $LiTi_2(PS_4)_3$ in the space group *Ccc*2.** Top: using SR-XPD data, λ = 0.50000 Å; bottom: using the NPD data, λ = 2.53612 Å. The blue marks show Bragg positions.

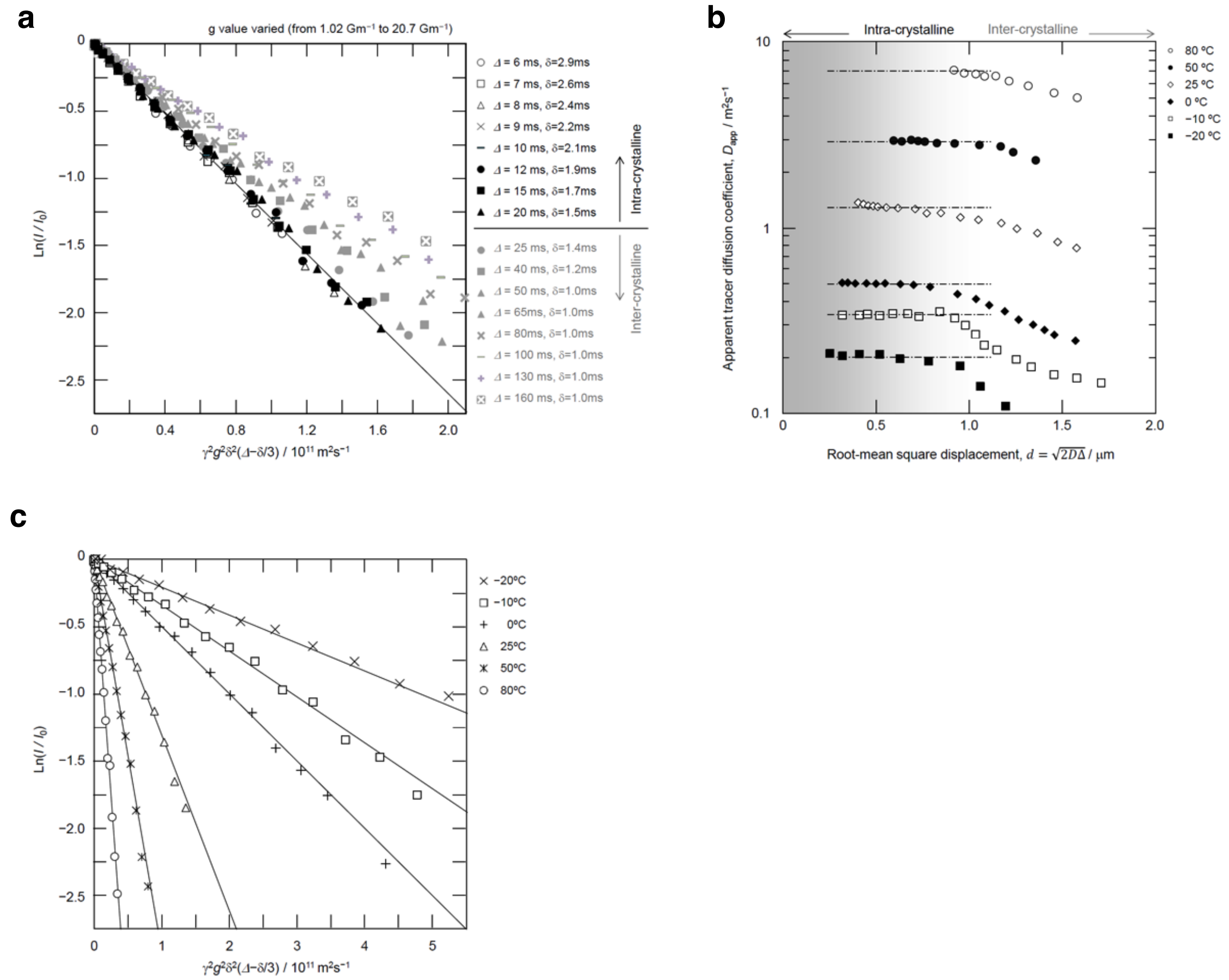

**Fig. S3.| $^{7}$Li PFG NMR spectrum of LTPS.** Echo attenuation curves measured at 25 °C with *Δ* varying from 6 ms to 160 ms (**a**). Root-mean square displacement dependence of apparent tracer diffusion coefficient (**b**). Selected echo attenuation curves at different temperatures from −20 °C to 80 °C (**c**).

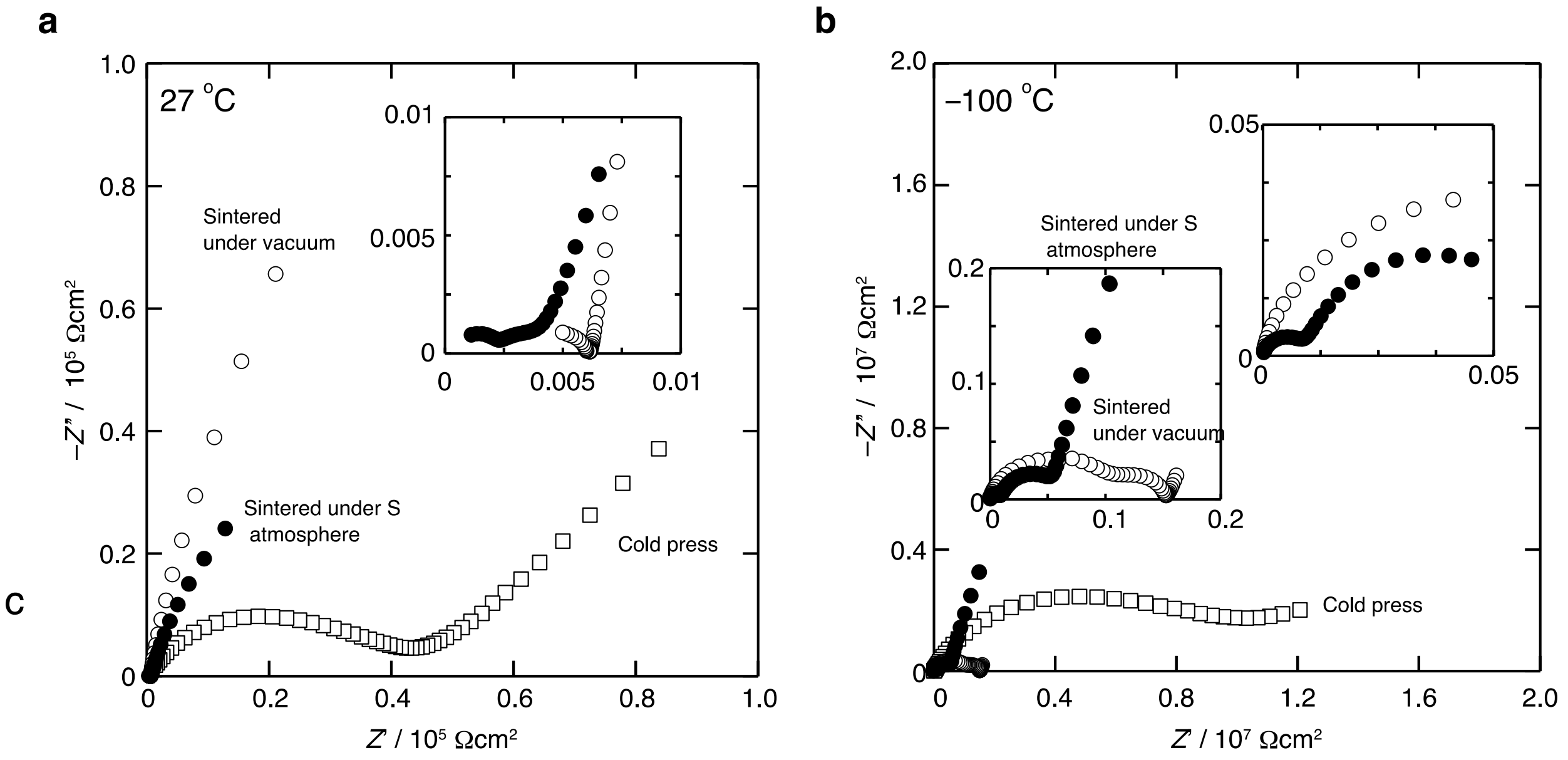

**Fig. S4.| Effect of sintering condition on impedance.** Nyquist plots of $LiTi_2(PS_4)_3$ pellets with different sintering conditions are shown at 27 °C (**a**) and at −100 °C (**b**).

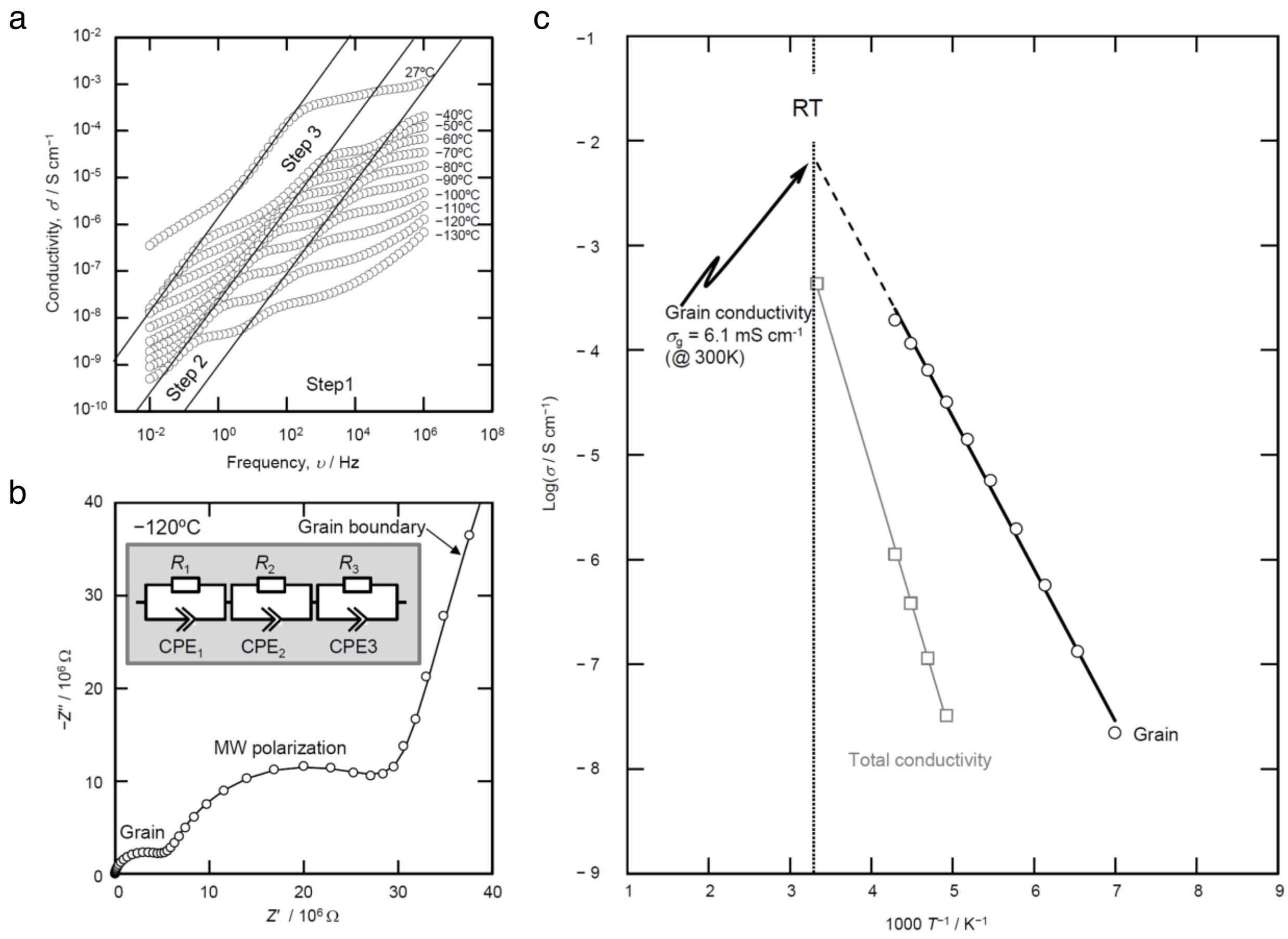

**Fig. S5| Impedance spectroscopy data of $LiTi_2(PS_4)_3$ pellet.** (**a**) Frequency dependence of the real part of conductivity $\sigma'$ at various temperatures. (**b**) Nyquist plot of the impedance at −120 ºC. Inset shows the equivalent circuit used for analysis. (**c**) Temperature dependence of electric total and grain conductivity.

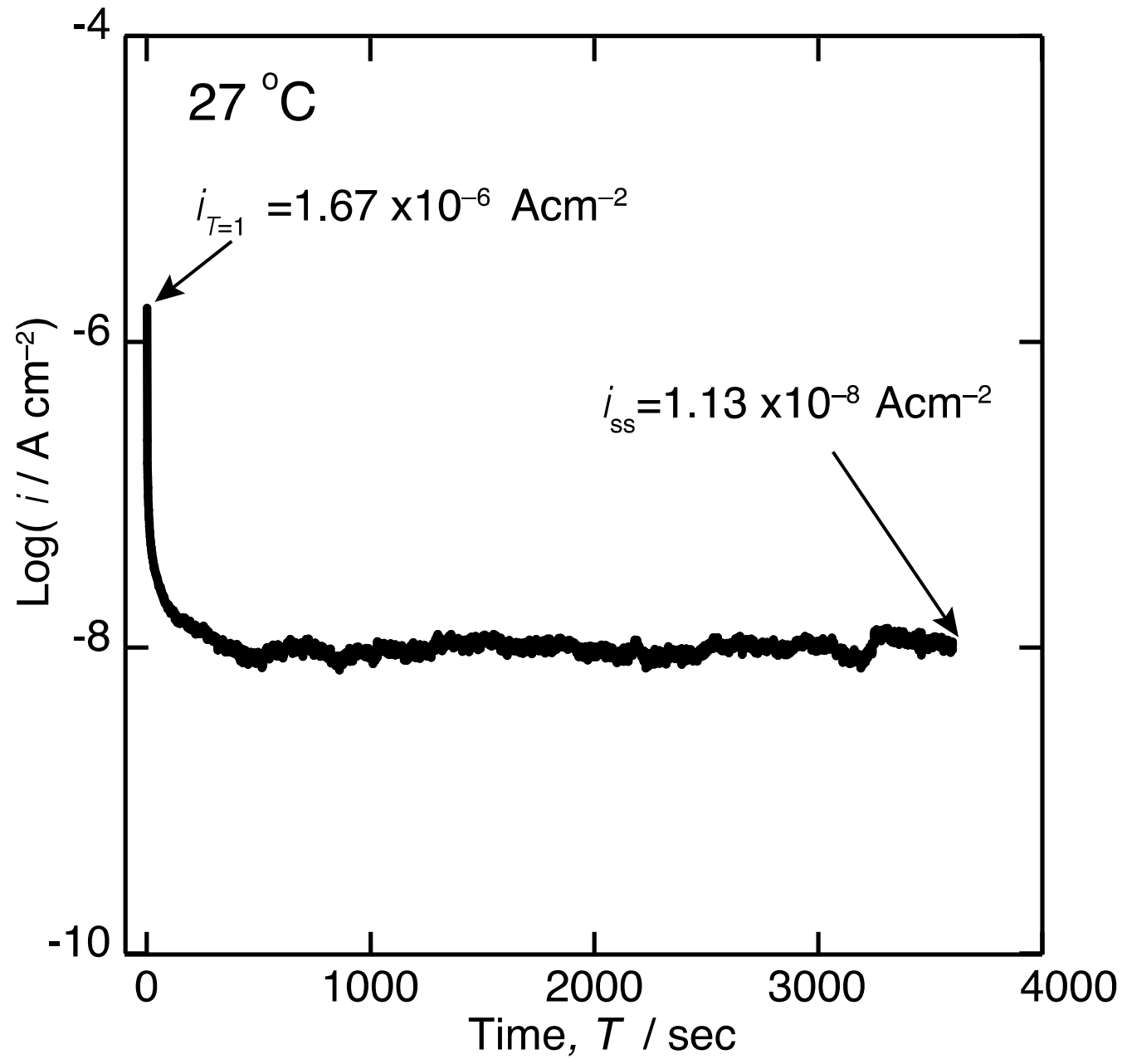

**Fig. S6| DC polarization measurement using ion blocking cell.** The transient current was measured until it reaches to steady state. The steady state current Iss was used for electronic conductivity estimation with the equation of $\sigma_e = (I_{ss} / V_{DC})*(L / S)$ where $\sigma_e$, $V_{DC}$, $L$ and $S$ are electronic conductivity, DC potential step, thickness and cross sectional area of the pellet, respectively.

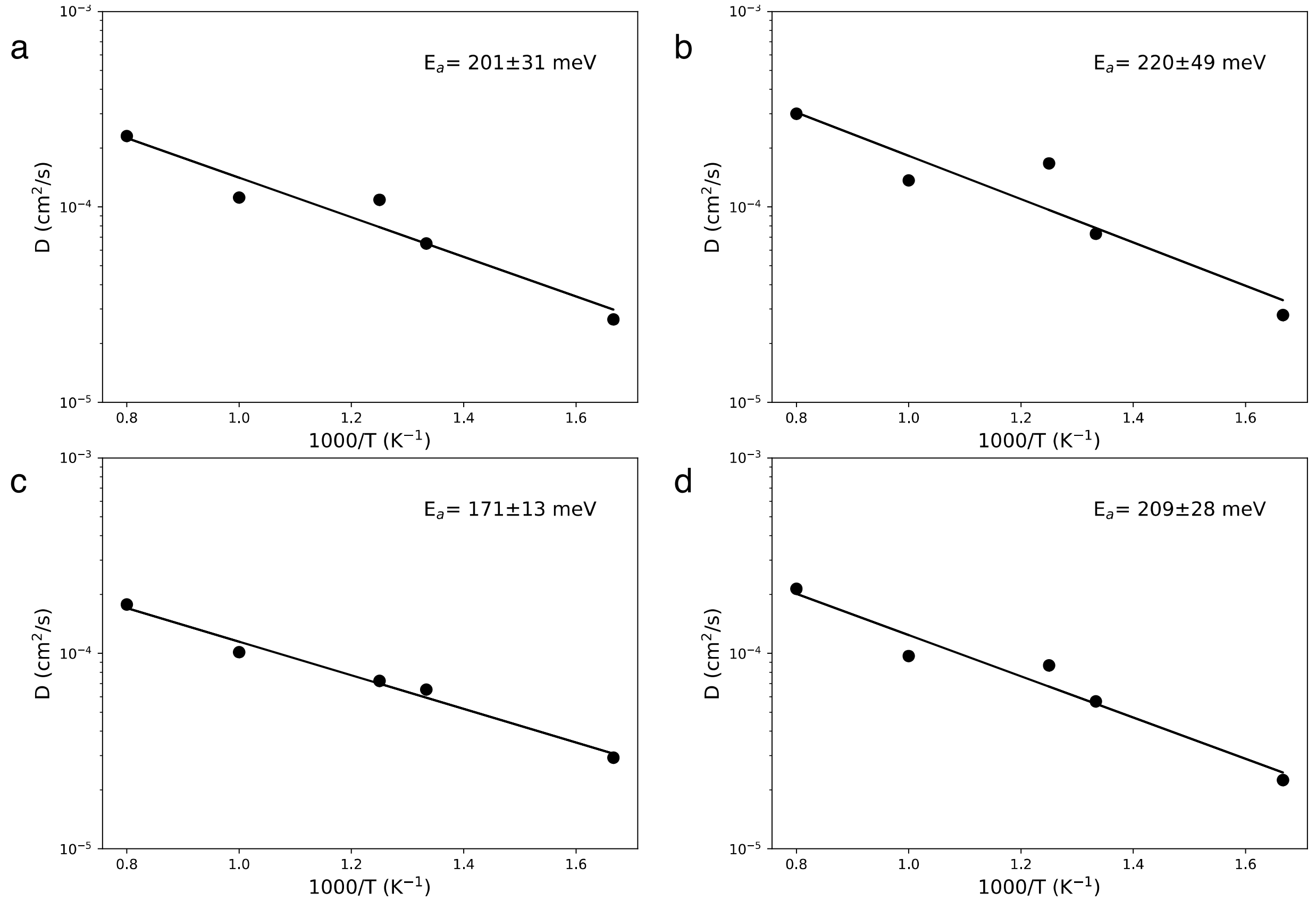

**Fig. S7| Diffusion coefficient vs temperature in $LiZr_2(PS_4)_3$ from AIMD.** Total diffusion coefficient (**a**) and components along the *a* axis (**b**), and in the *b*-*c* plane (**c** and **d**). The $LiZr_2(PS_4)_3$ diffusion properties are similar to LTPS with an activation energy of 201 meV.

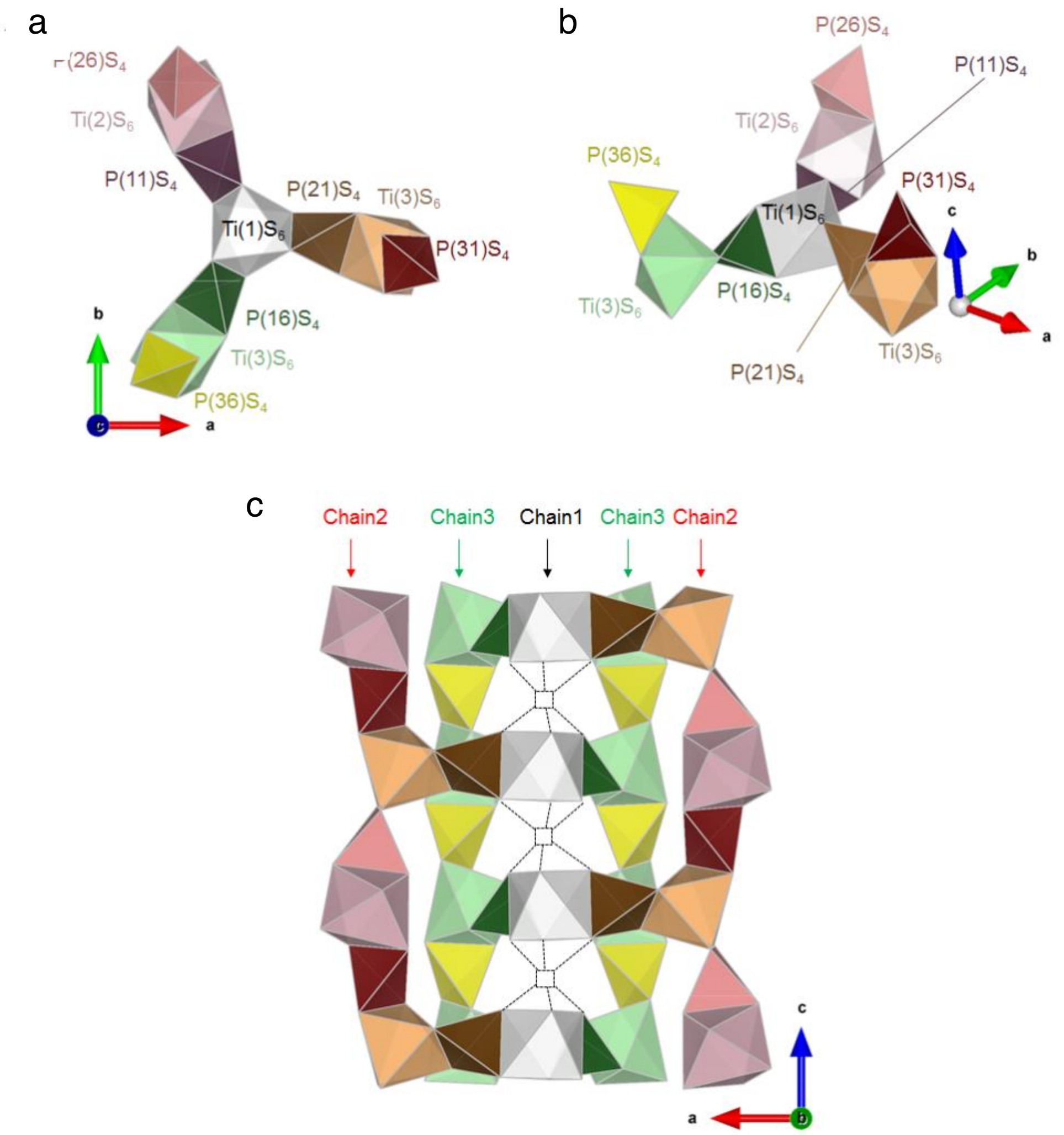

**Fig. S8| Ti-P-S framework structure of $LiTi_2(PS_4)_3$.** (**a** and **b**) $Ti_4P_6S_{30}$ polyhedral unit. Ti(1) is on the three-fold-like axis and the polyhedral block of $Ti_4P_6S_{30}$ is generated along the three-fold-like axis (**c**) 1D chains along *c* axis. The polyhedral is regenerated every *c*/2 along the glide plane perpendicular to *b* axis. The 3D Ti-P-S framework consists in three different 1D chains along *c* axis [chain1: face-sharing -Ti(1)$S_6$-(*Vacancy*)$S_6$-, chain2: edge-sharing -Ti(2)$S_6$-P(31)$S_4$-Ti(4)$S_6$-P(26)$S_4$-, chain3: -Ti(3)$S_6$-P(36)$S_4$-] and those chains are connected by $PS_4$ tetrahedra sharing edge with Ti(1)$S_6$ octahedra

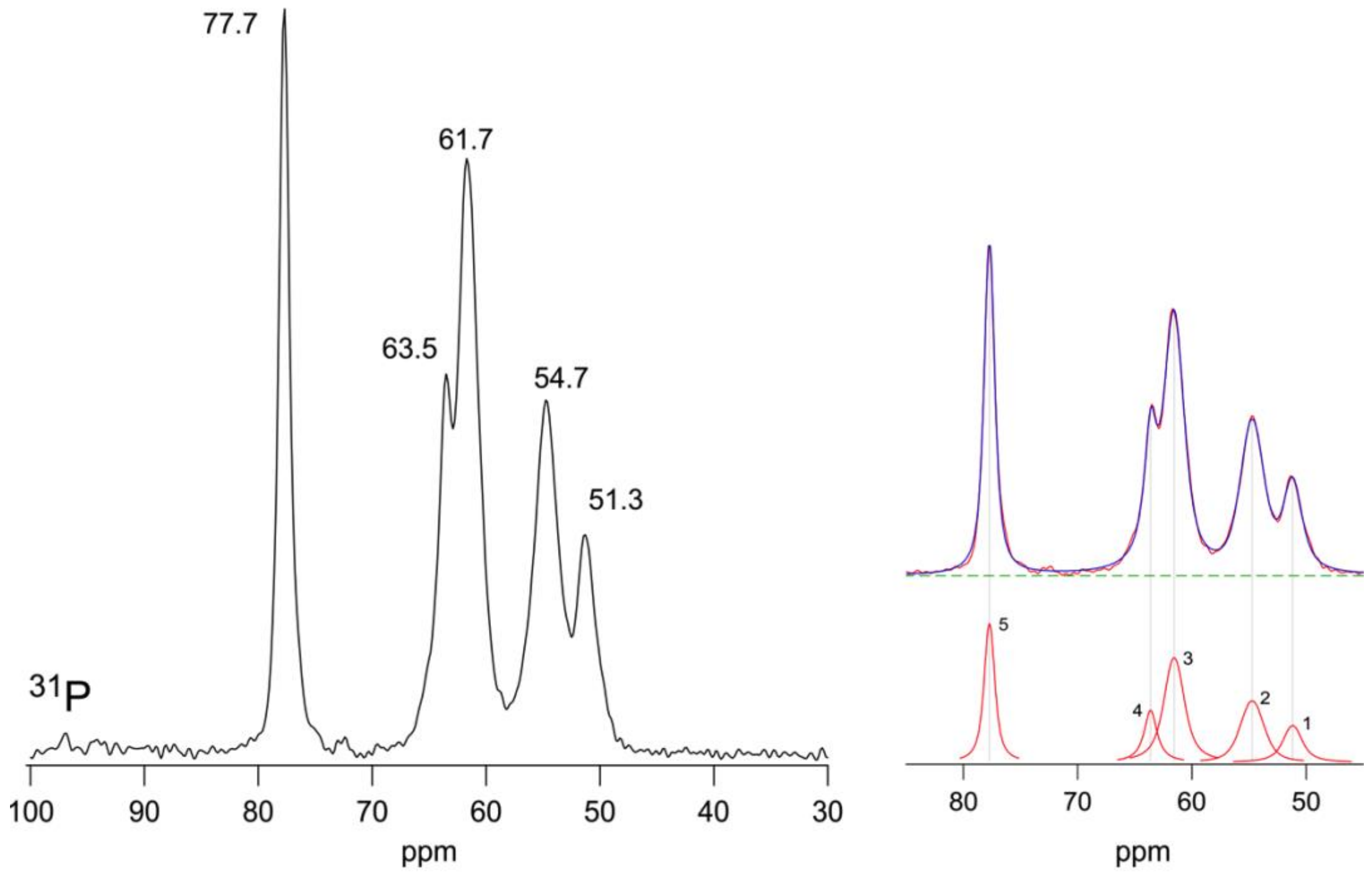

**Fig. S9| [31]P MAS NMR of LTPS** Left: $^{31}$P MAS NMR spectrum (202.4 MHz, 25 kHz). Right: deconvolution of the spectrum with appropriate Voigt functions. The spectrum reveals at least 5 magnetically distinct NMR lines, which is in good agreement with the crystal structure of LTPS with its 6 crystallographically different P positions. A rotation speed of 25 kHz limited the resolution of our experiment. Between 50 and 70 ppm an additional $^{31}$P signal might show up at higher spinning speeds. Narrow NMR lines point to a well crystalline sample. $^{31}$P MAS NMR does not reveal any large amounts of foreign phases such as the highly conductive thio-phosphate glass, $x$$Li_2S$-(1−$x$)$P_2S_5$. Chemical shifts of the latter are characterized by values larger than 80 ppm ($PS_4^{3-}$; 86 ppm, $P_2S_4^{4-}$; 91 ppm, $P_2S_6^{4-}$; 109 ppm).[64, 65] Furthermore, the line widths of glassy phases are generally larger than those of crystalline compounds. Thus, only tiny or no amounts of glassy phases are expected in our powder sample. Their impact on the bulk Li ion dynamics as probed through PFG NMR turns out to be negligible. Long-range as well as local ion dynamics, seen by $^{7}$Li PFG and spin-lattice relaxation NMR, is solely due to LTPS.

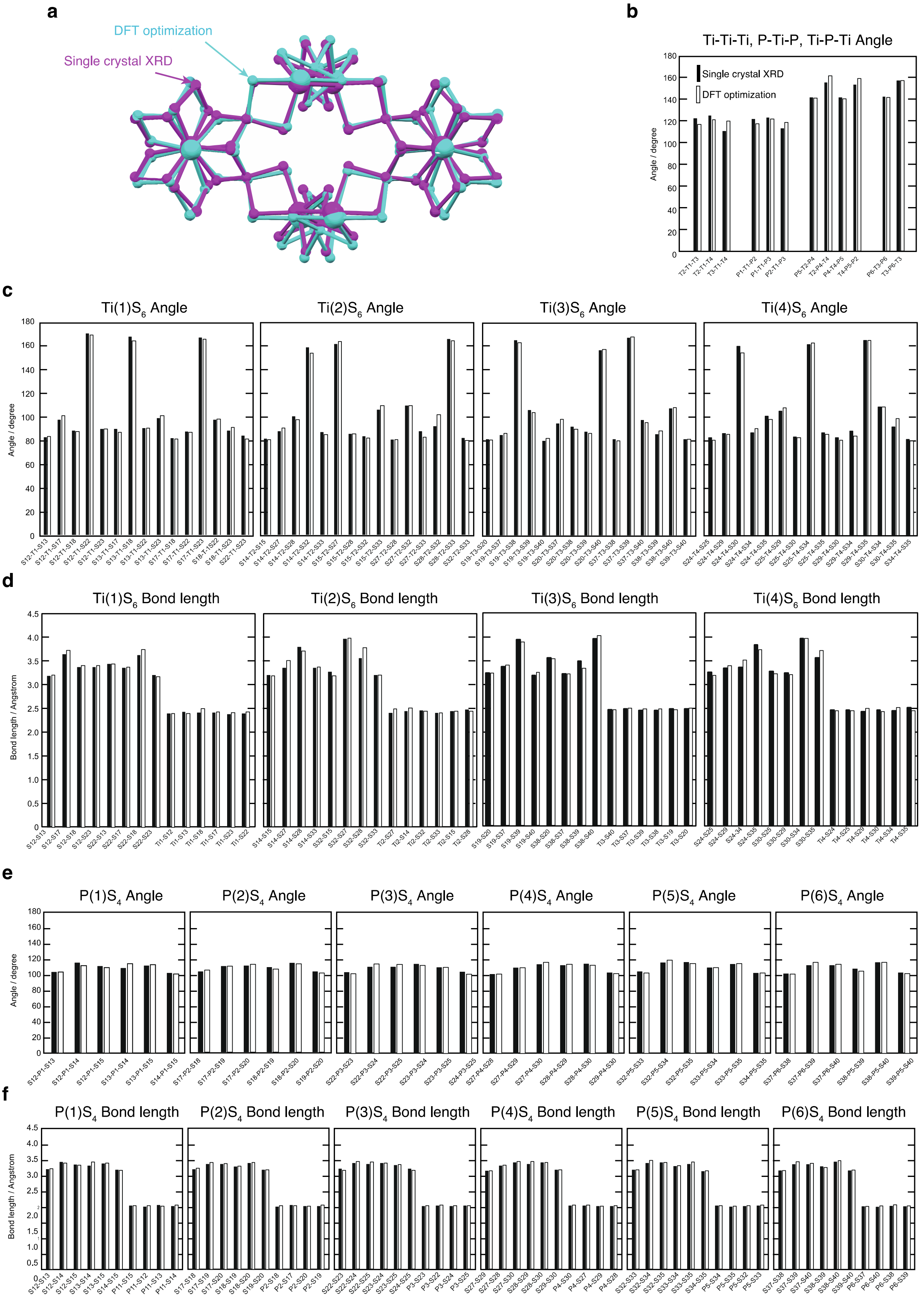

**Fig. S10. Structure comparison between DFT optimized one (*P*1) from *P*6*cc* symmetry and experimentally solved one (*Ccc*2) by single crystal X-ray diffraction. a,** Overlay of structures obtained by experiment and DFT optimization. The structure is shown along [001]. **b,** Angle of framework skeleton. **c,d,** Angle and bond length of $TiS_6$ octahedra respectively. **c,d,** Angle and bond length of $PS_4$ tetrahedra, respectively.

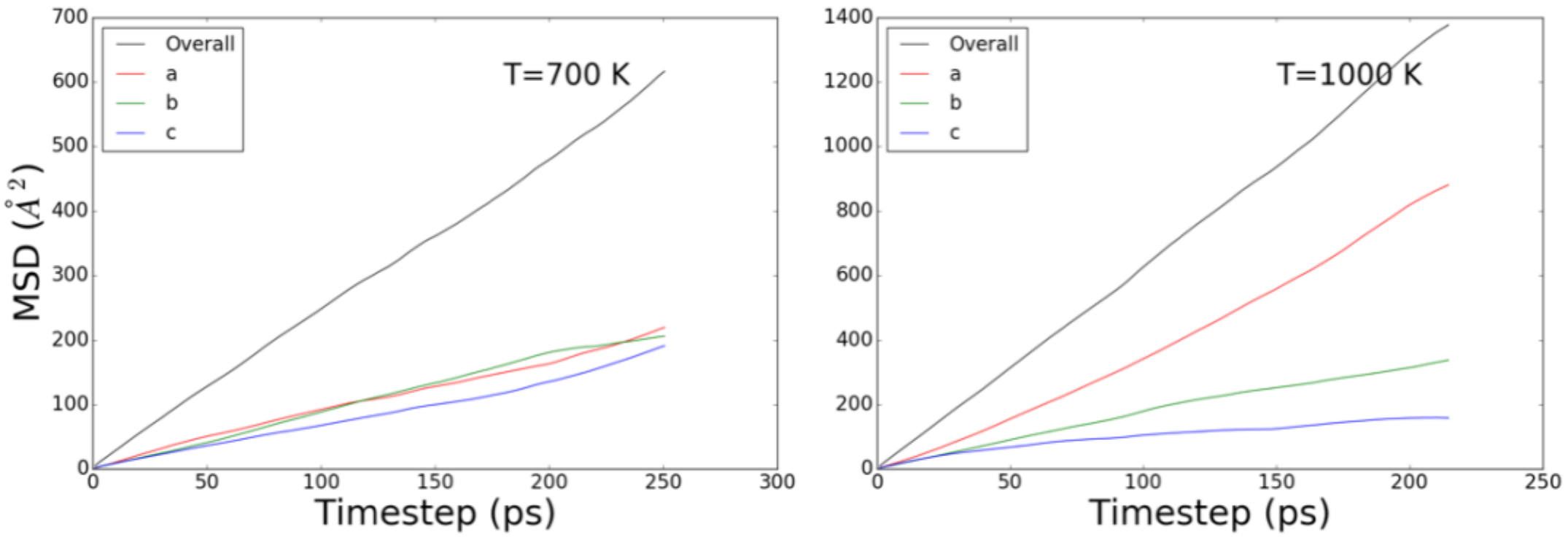

**Fig. S11| Mean square displacement for lithium atoms (MSD) as function of time for LTPS from AIMD simulations.** The black line gives the overall MSD and the red, green and blue line give the MSD along the three a,b,c axis in the hexagonal settings. The MSD at a temperature of 700K (left) and 1000K (right) are given.

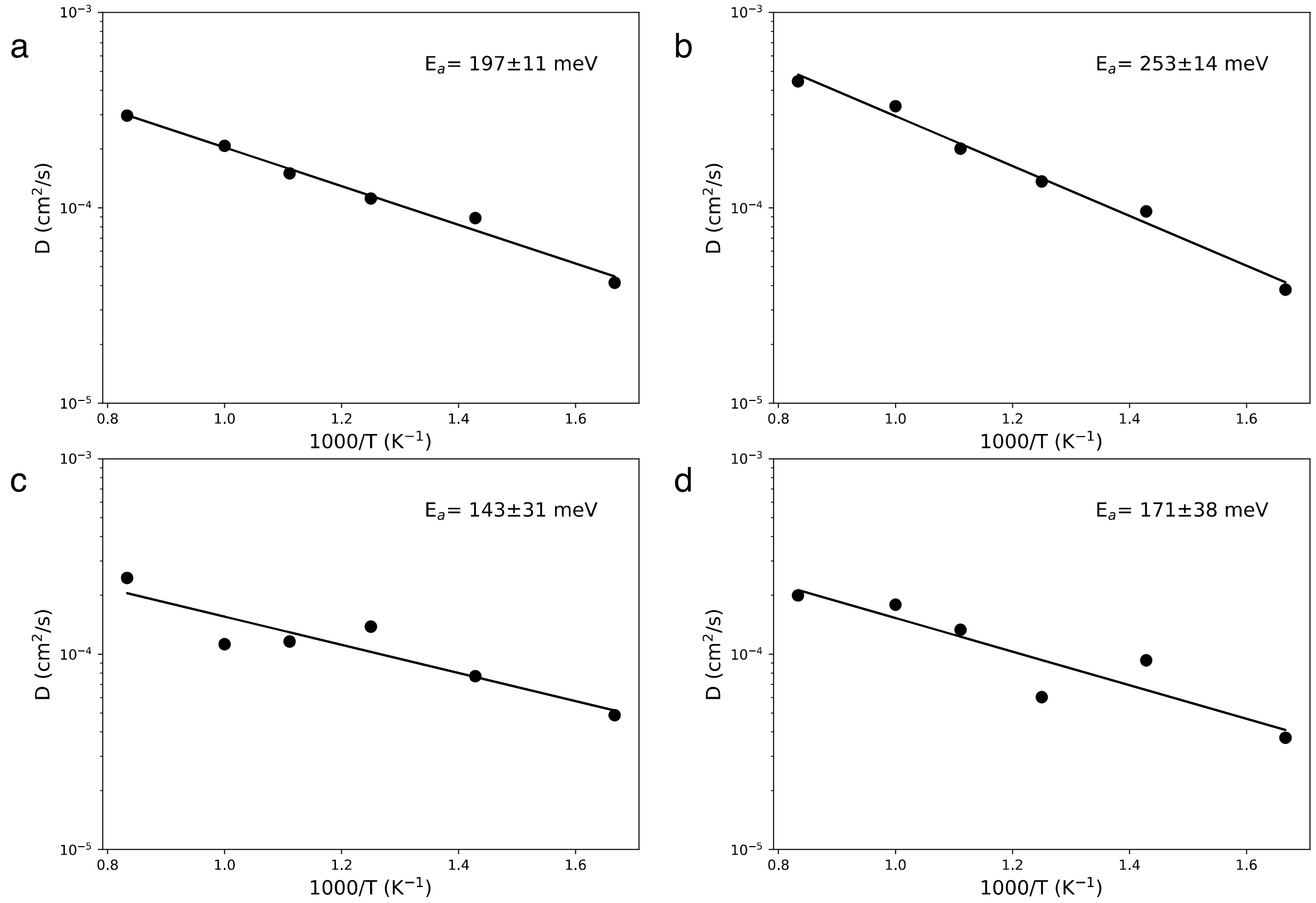

**Fig. S12| Lithium tracer diffusion vs temperature from AIMD.** Total diffusion coefficient (**a**) and components along the [100] (**b**), [001] (**c**), and [010] (**d**) and directions in the hexagonal settings.

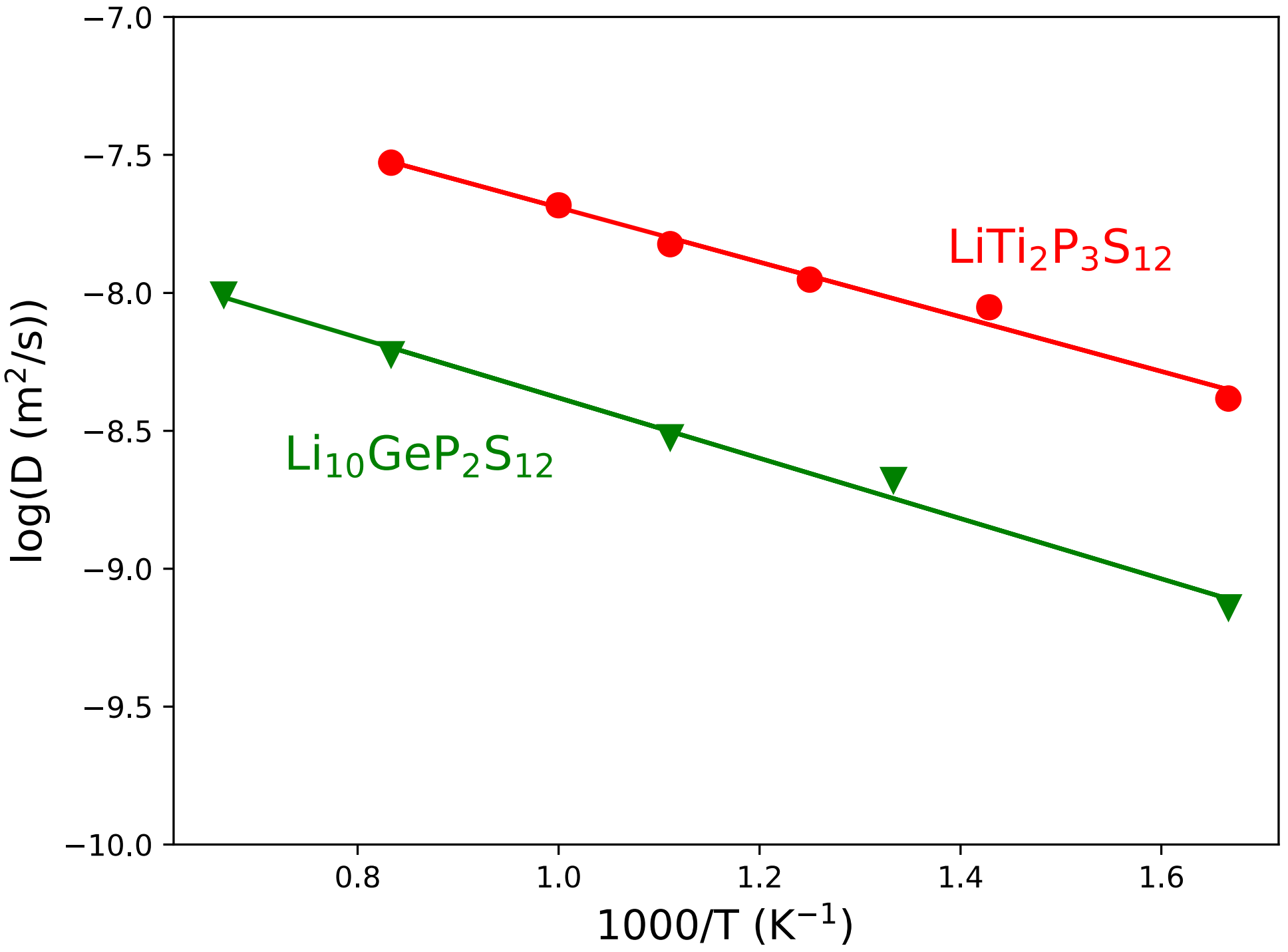

Fig. S13| Comparison between computed lithium tracer diffusion coefficients from AIMD in LGPS and LTPS

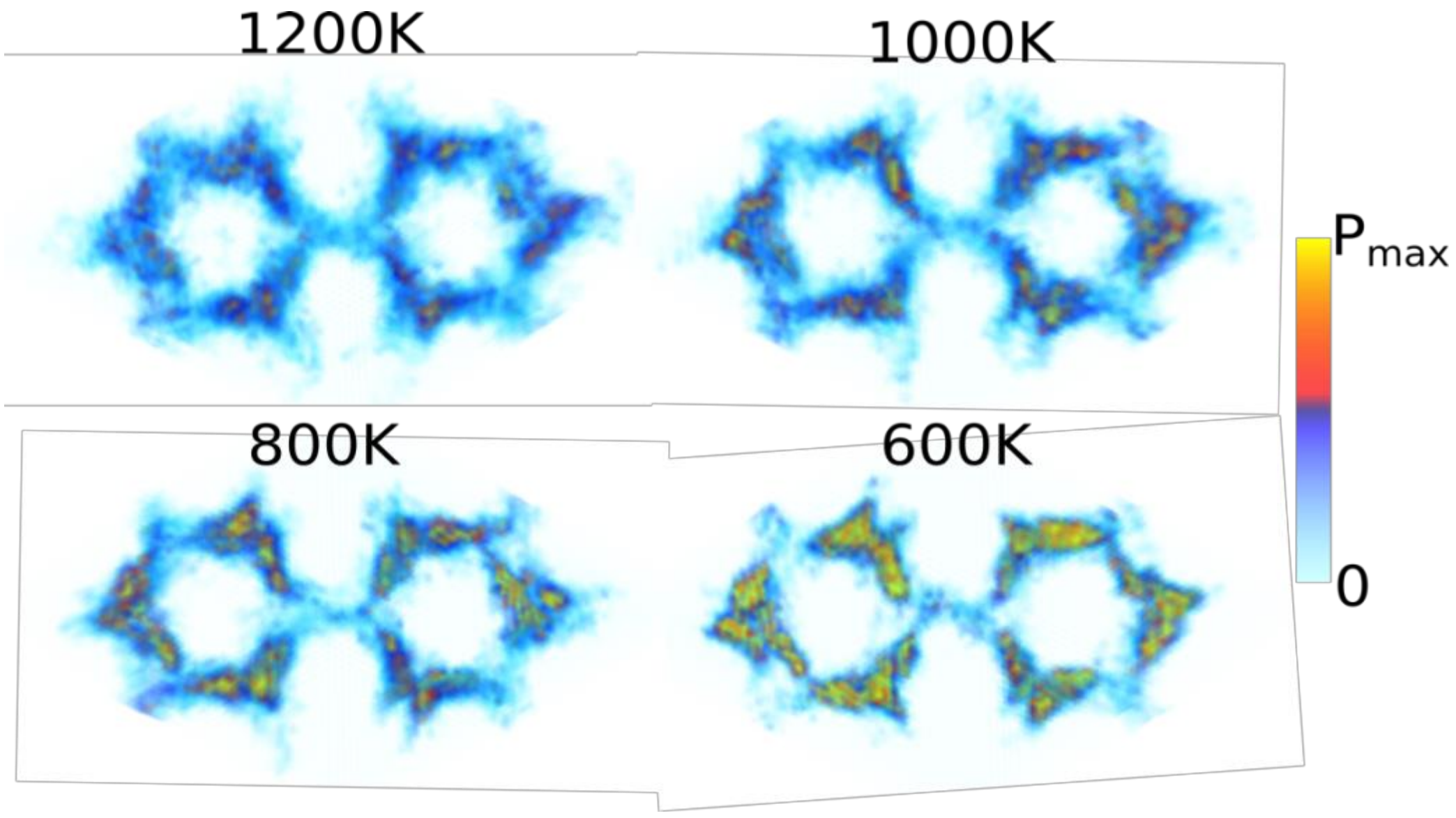

Fig. S14| Top view of probability density of lithium presence in LTPS at different temperature from AIMD computations (600K, 800K, 1000K, 1200K).

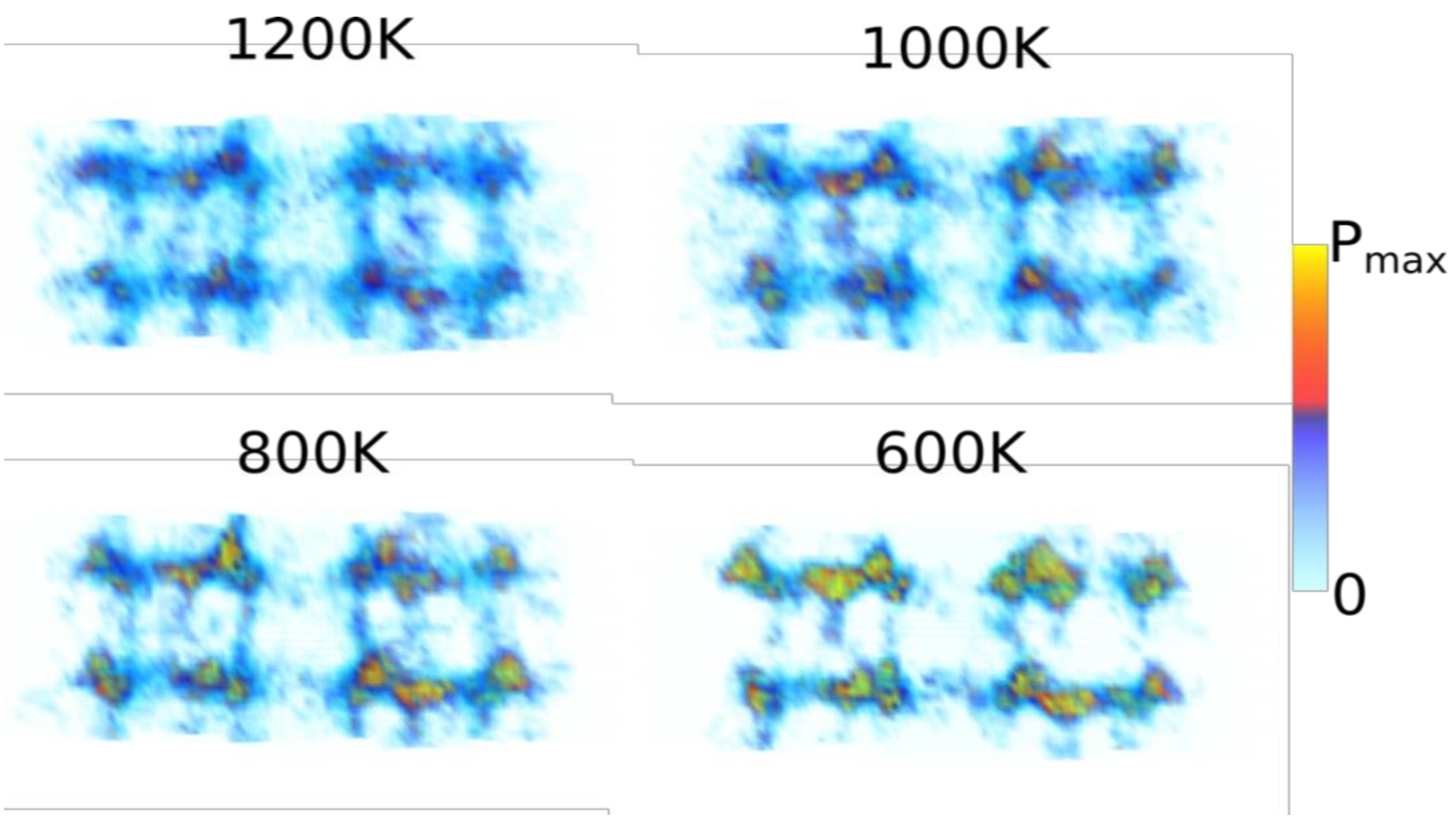

Fig. S15| Side view of probability density of lithium presence in LTPS at different temperature from AIMD computations (600K, 800K, 1000K, 1200K).

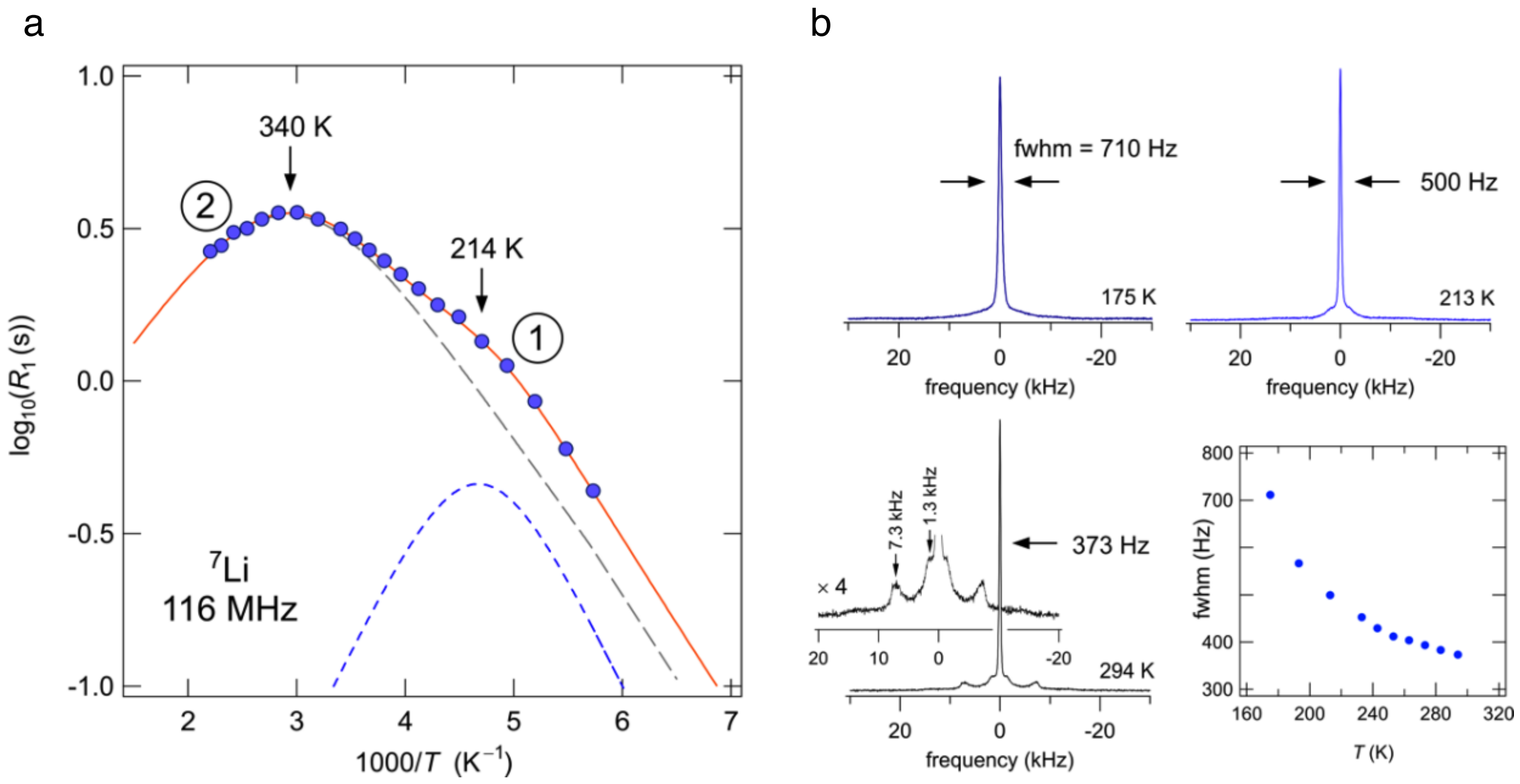

**Fig. S16| $^{7}$Li NMR spin-lattice relaxation of LTPS.** (**a**) The spin-lattice relaxation rate ($R_1$) vs temperature. $R_1$ pass through two relaxation rate peaks revealing complex Li self-diffusion behaviour in LTPS with at least two diffusion processes running in parallel or being stepwise activated with increasing temperature. At the peak maxima the corresponding jump rates are in the order of $7.3 \times 10^{8}$ s$^{\square 1}$. (**b**) The corresponding $^{7}$Li NMR lines recorded under static conditions reveal pronounced motional narrowing due to extremely fast Li hopping processes in LTPS. A sharp quadrupole powder pattern shows up at ambient temperature. At this temperature, the NMR central line is fully averaged due to extremely fast ion hopping indicating fully averaged line shapes.

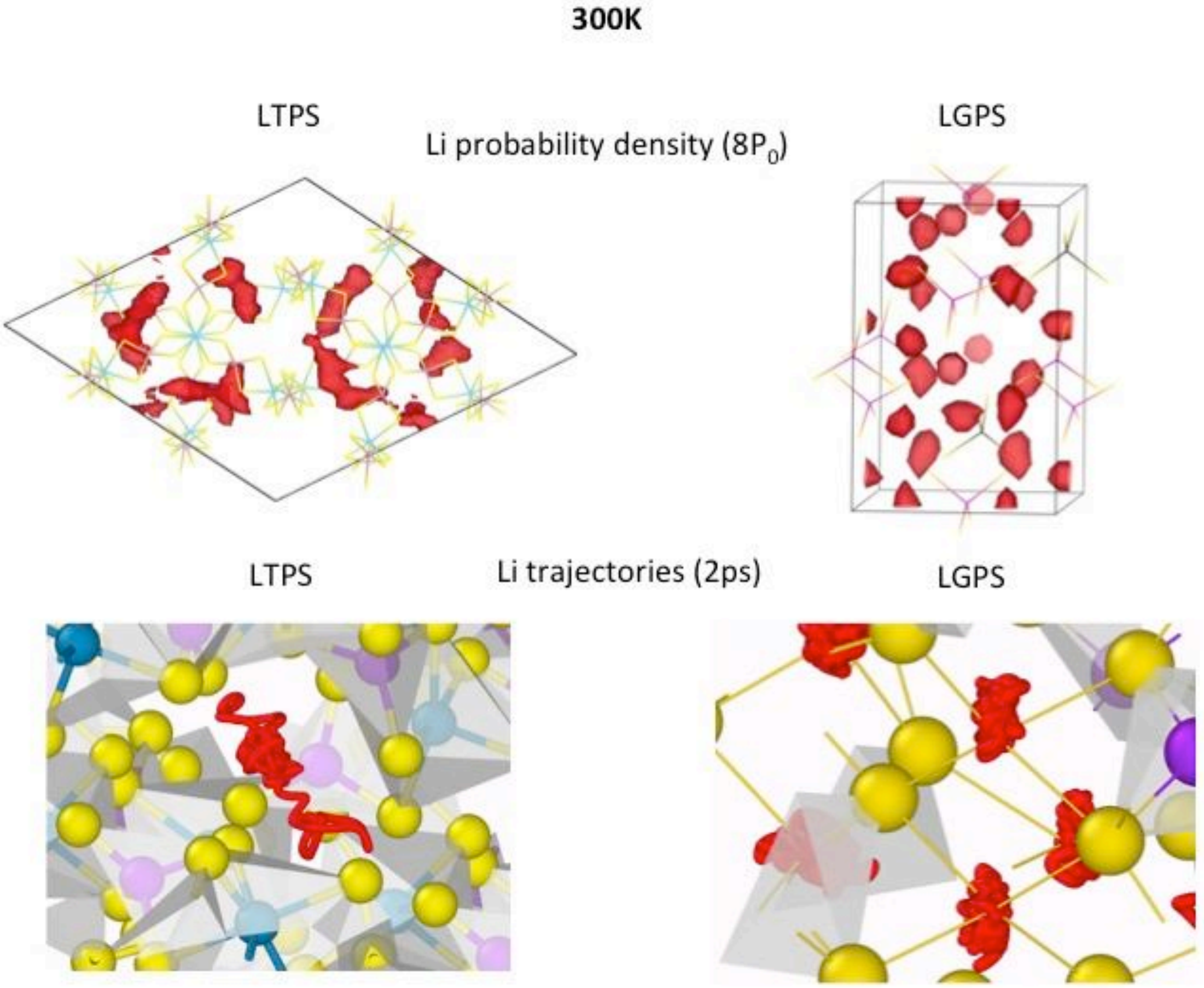

**Fig. S17| Lithium probability density in LTPS and LGPS from AIMD at 300K.** Probability densities are obtained from 40ps runs at 300K. The probability threshold is set to $8P_0$ where $P_0$ is the average lithium probability. The much larger potato-shaped stable sites in LTPS are clearly seen in contrast to the tetrahedral well-localized occupations in LGPS. Representative 2ps lithium trajectories are also shown in LTPS and LGPS at 300K. The lithium are vibrating around the center of a tetrahedral while in LTPS lithium travels among several distorted tetrahedra.

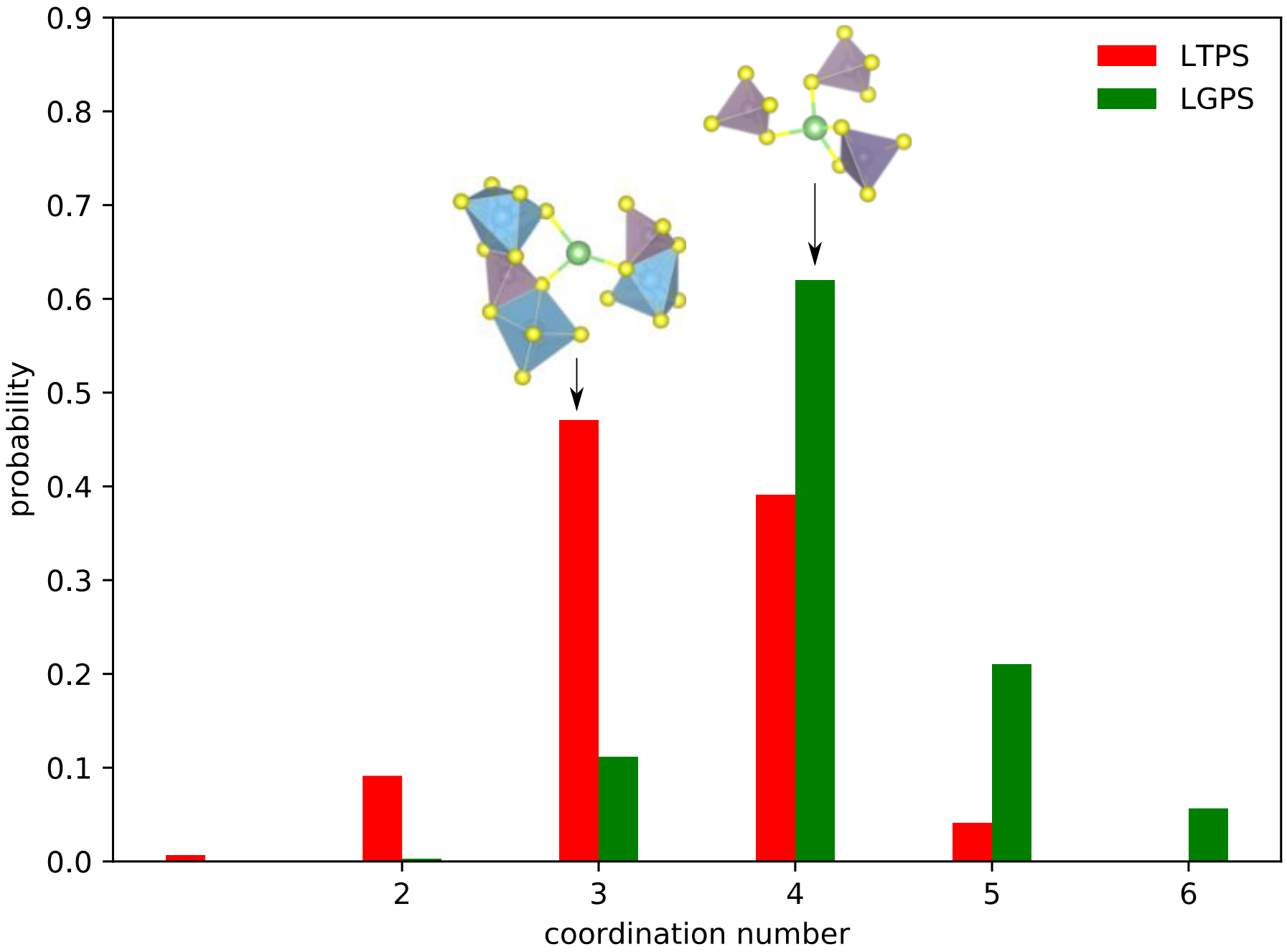

**Fig. S18| Instantaneous coordination number in LTPS and LGPS.** The instantaneous coordination number is defined as the number of sulfur atoms at a distance shorter than the Li-S radial function distribution first minimum. A cutoff distance of 3.0 Å (Li-S typical bond is 2.5 A), for both LTPS and LGPS, is obtained from the analysis of Li-S radial distribution functions computed from MD simulations. The values were obtained from 10 ps AIMD on LTPS and LGPS at 600K. We see clearly a smaller coordination number for LTPS than for LGPS. The coordination of LTPS is mainly 3-fold while LGPS is 4-fold (tetrahedral). We added a representative picture of lithium and its coordination environment on the Figure.

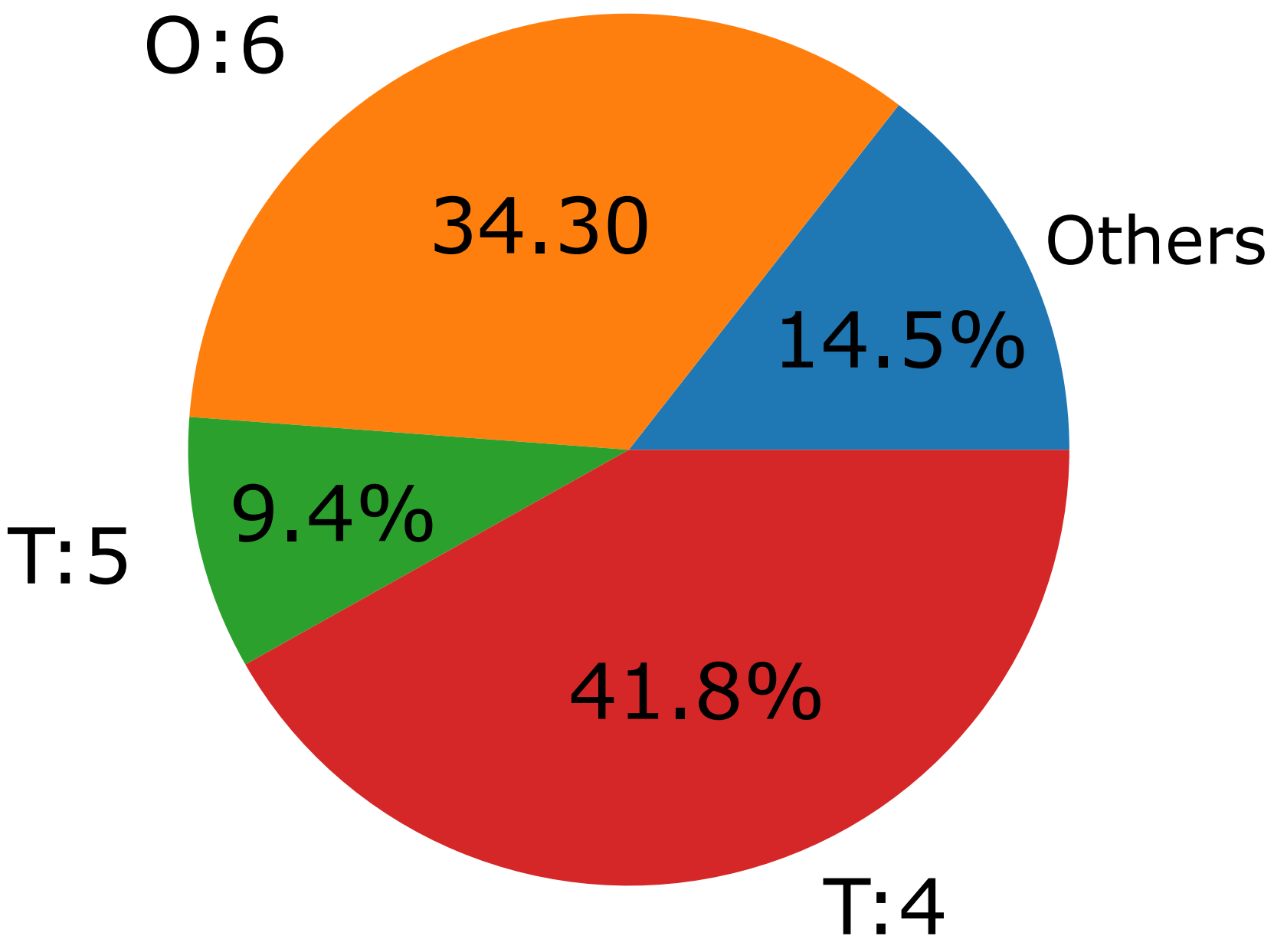

**Fig. S19| Distribution of local environments of lithium in sulfides.** Structures present in the Materials Project database and originating from the Inorganic Crystal Structure Database were considered. T:4 refers to tetrahedral, O:6 to octahedral, T:5 to trigonal bipyramidal and "Others" to any other local environment.

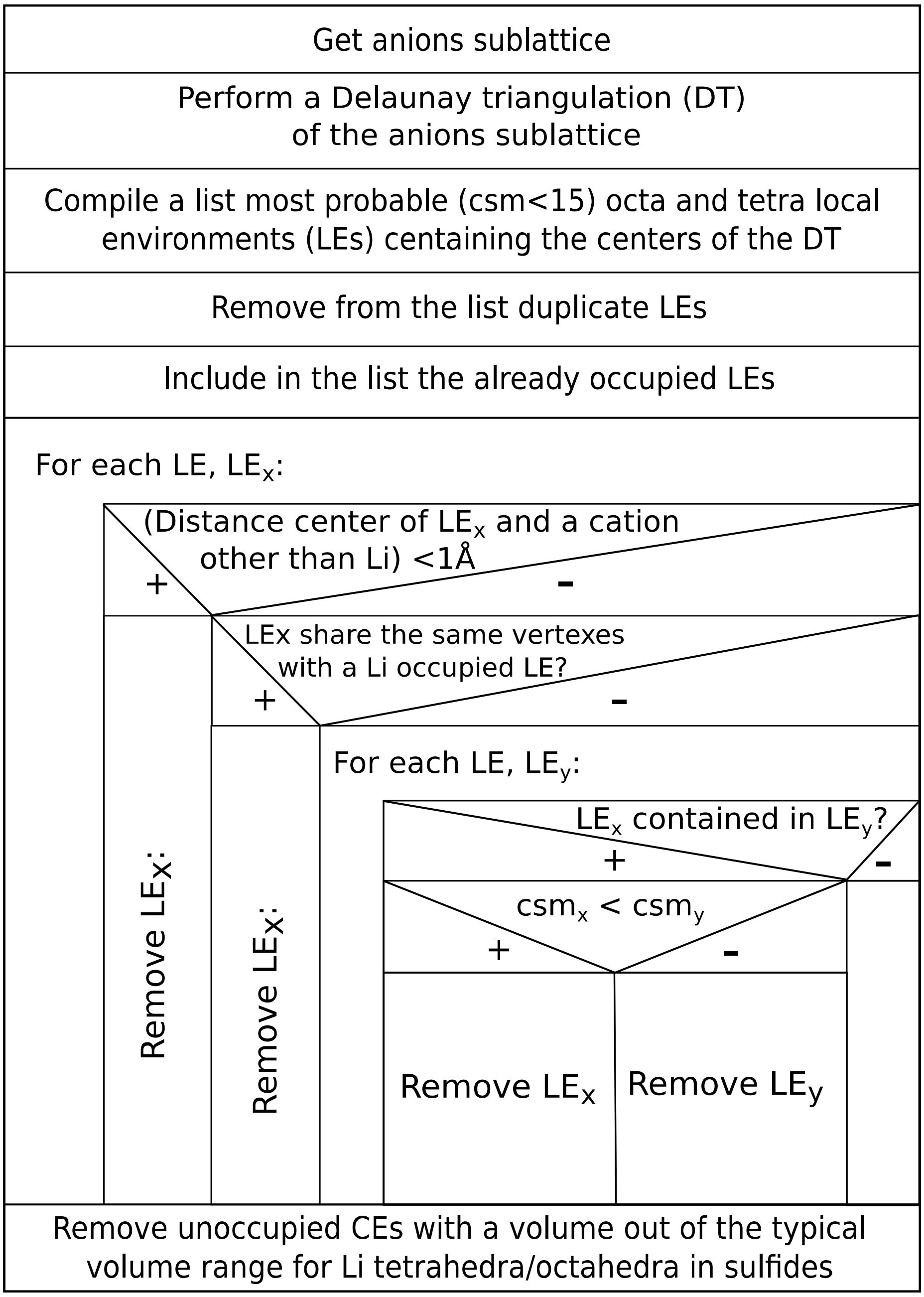

Fig. S20| NS-diagram representing the algorithm designed to compile a list of occupied and unoccupied tetrahedral and octahedral local environment (LE) in LTPS and LGPS.

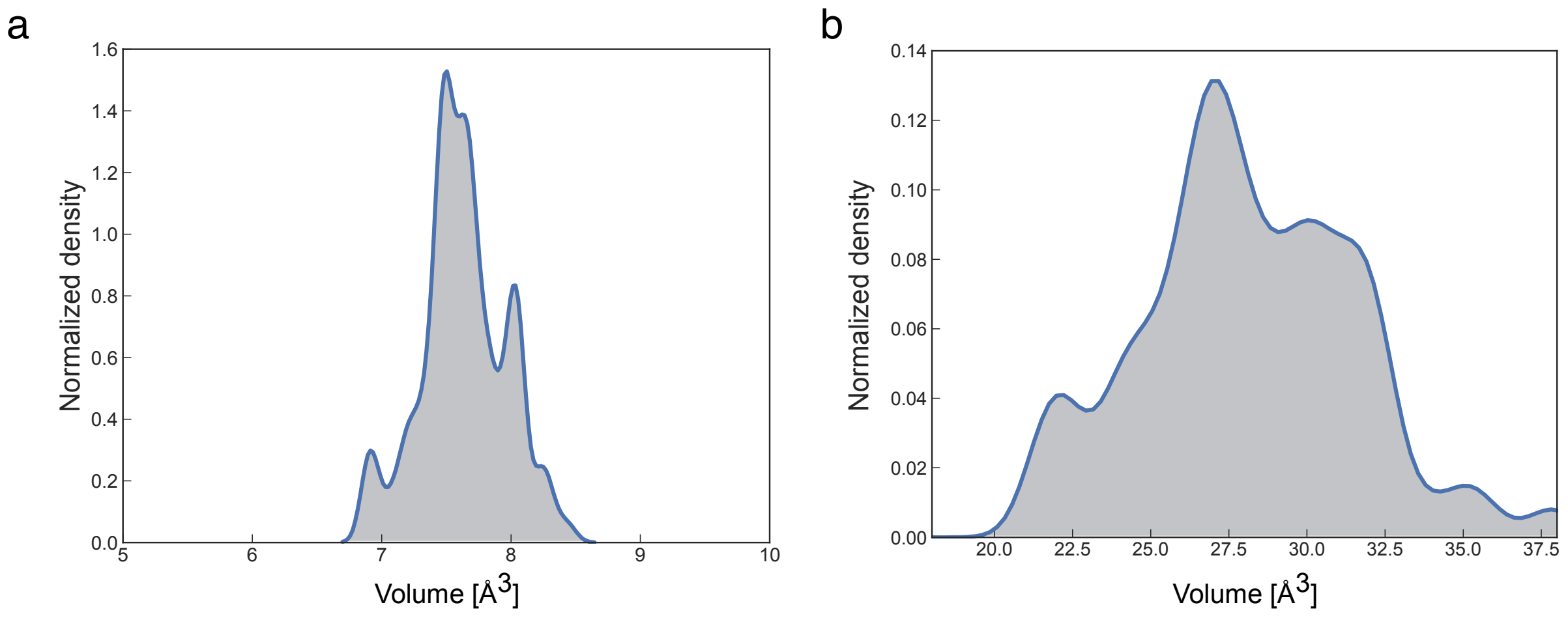

Fig. S21| Volume distribution for the Li occupied tetrahedral or octahedral local environments in stable sulfides in the Materials Project database. (a) tetrahedral and (b) octahedral.

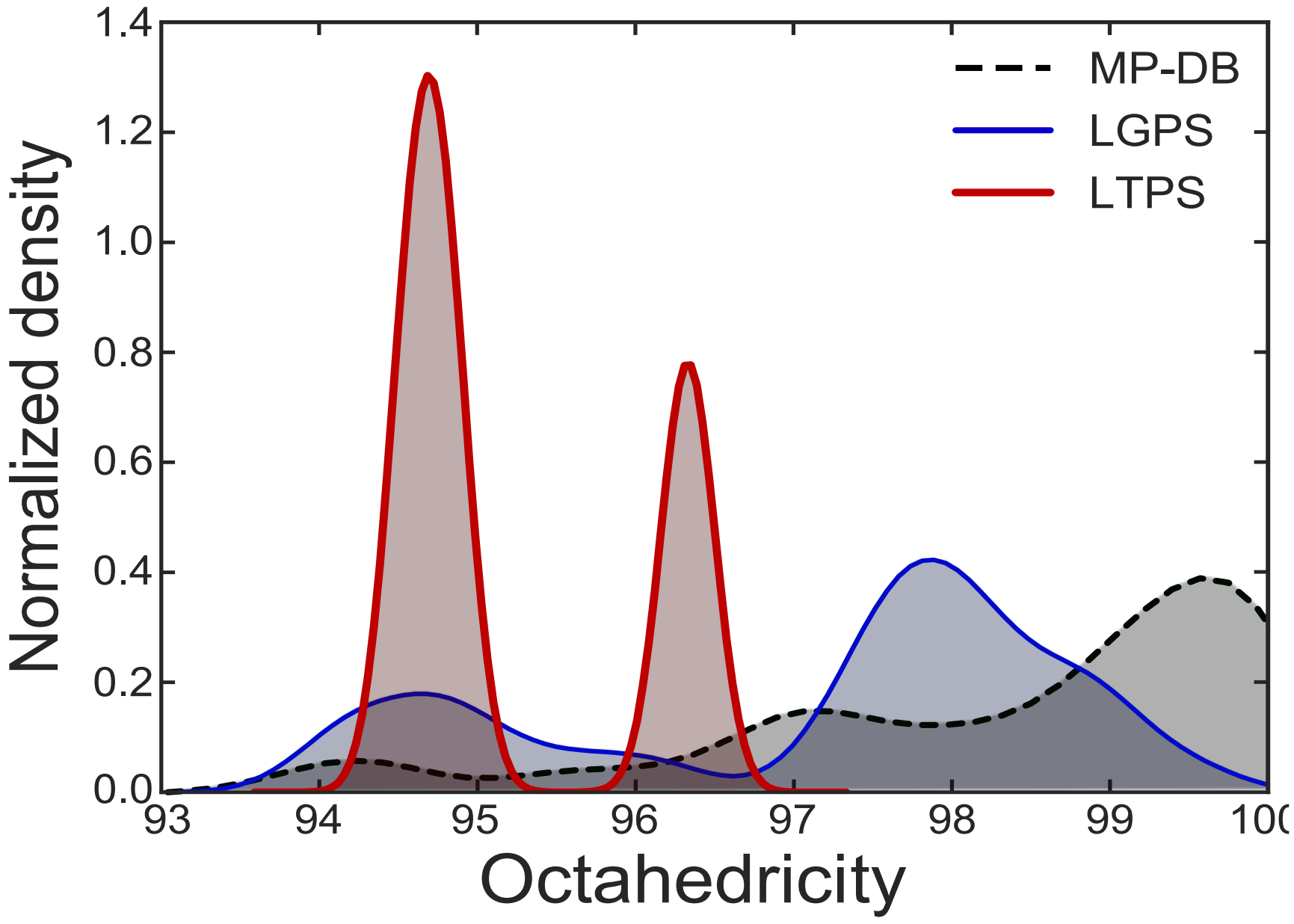

**Fig. S22| Distortion distribution of octahedral local environments.** In dashed black, octahedral site distortion distribution for the Li-occupied sites in stable sulfides in the Materials Project database. In red, octahedral sites distortion distribution for all (occupied and unoccupied) sites in LTPS. In blue, octahedral sites distortion distribution for all (occupied and unoccupied) sites in LGPS

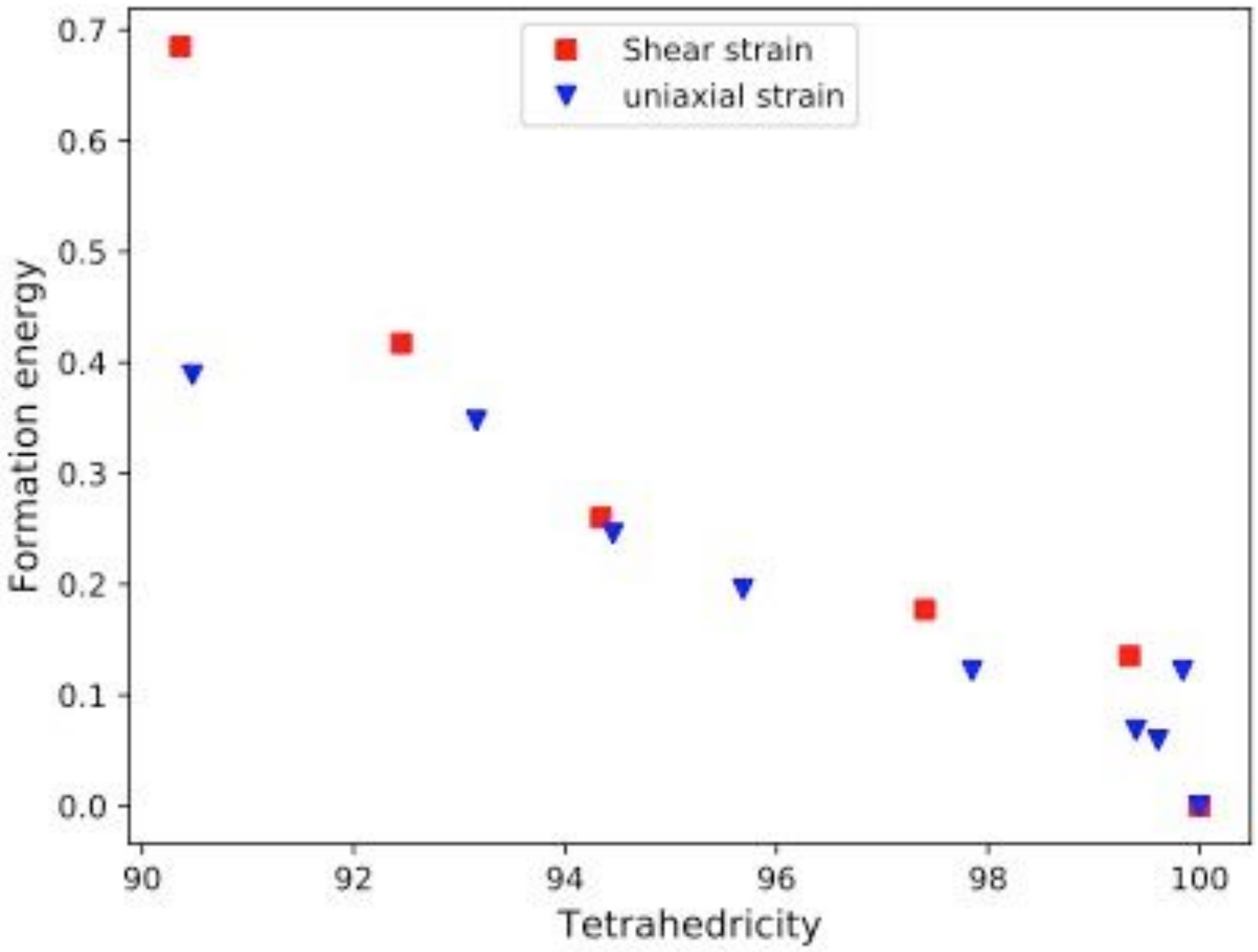

**Fig. S23| Effect of tetrahedricity on the energetics of an isolated $LiS_4$ environment.** We observe a clear relationship between the distortions which decrease tetrahedricity and a raise in the formation energy. This indicates that when deviating from the perfect tetrahedron, the occupation of the $S_4$ tetrahedron with lithium is less energetically favorable.

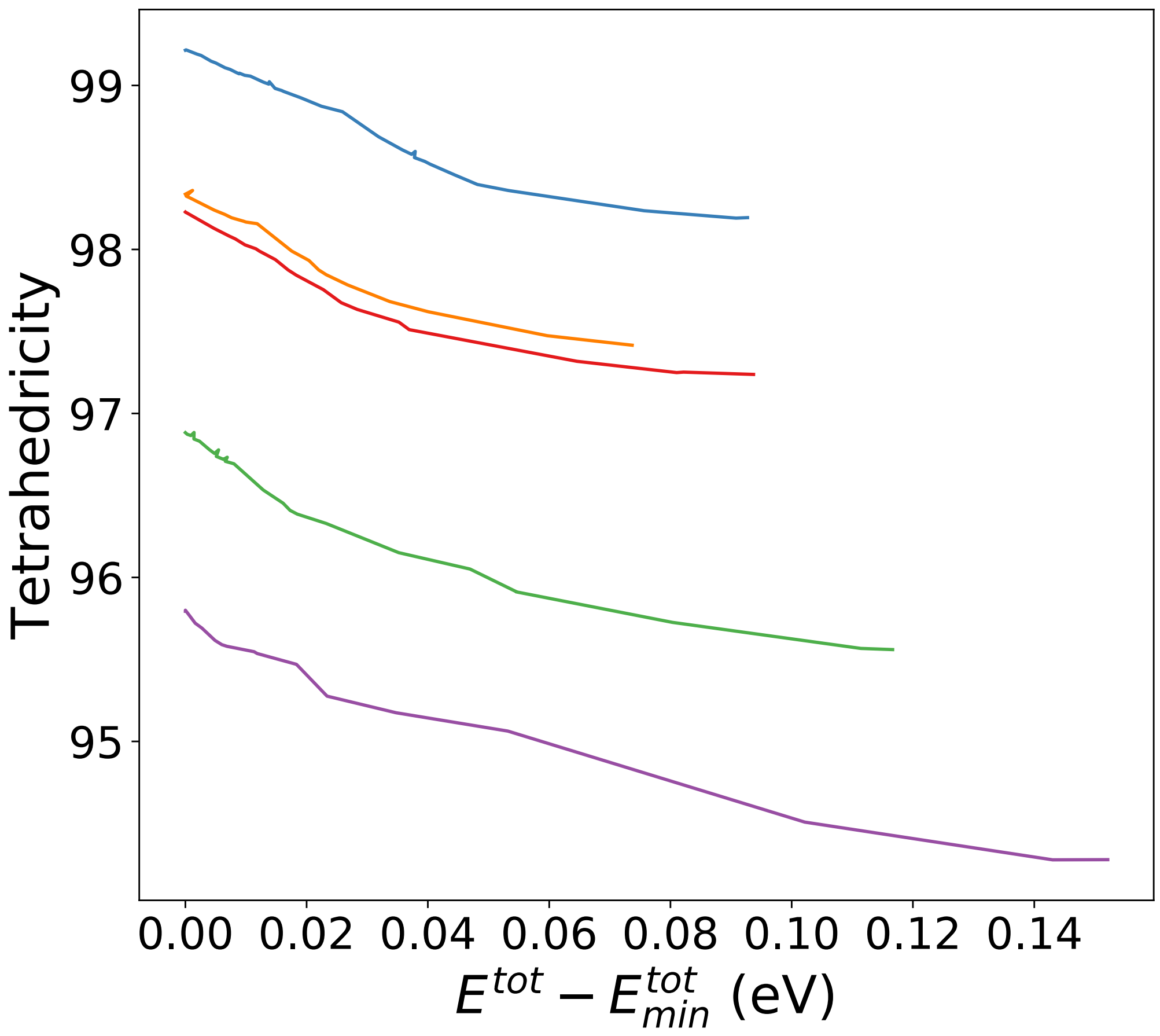

**Fig. S24| Tetrahedricity versus energy for various distorted starting S bcc structure with a Li in a tetrahedral site.** During the ionic relaxation, the energy tends to decrease (as expected from a structural relaxation) but this decrease in energy is always correlated with an increase in tetrahedricity. This shows that the energy is strongly related to the tetrahedricity of the lithium local environment.

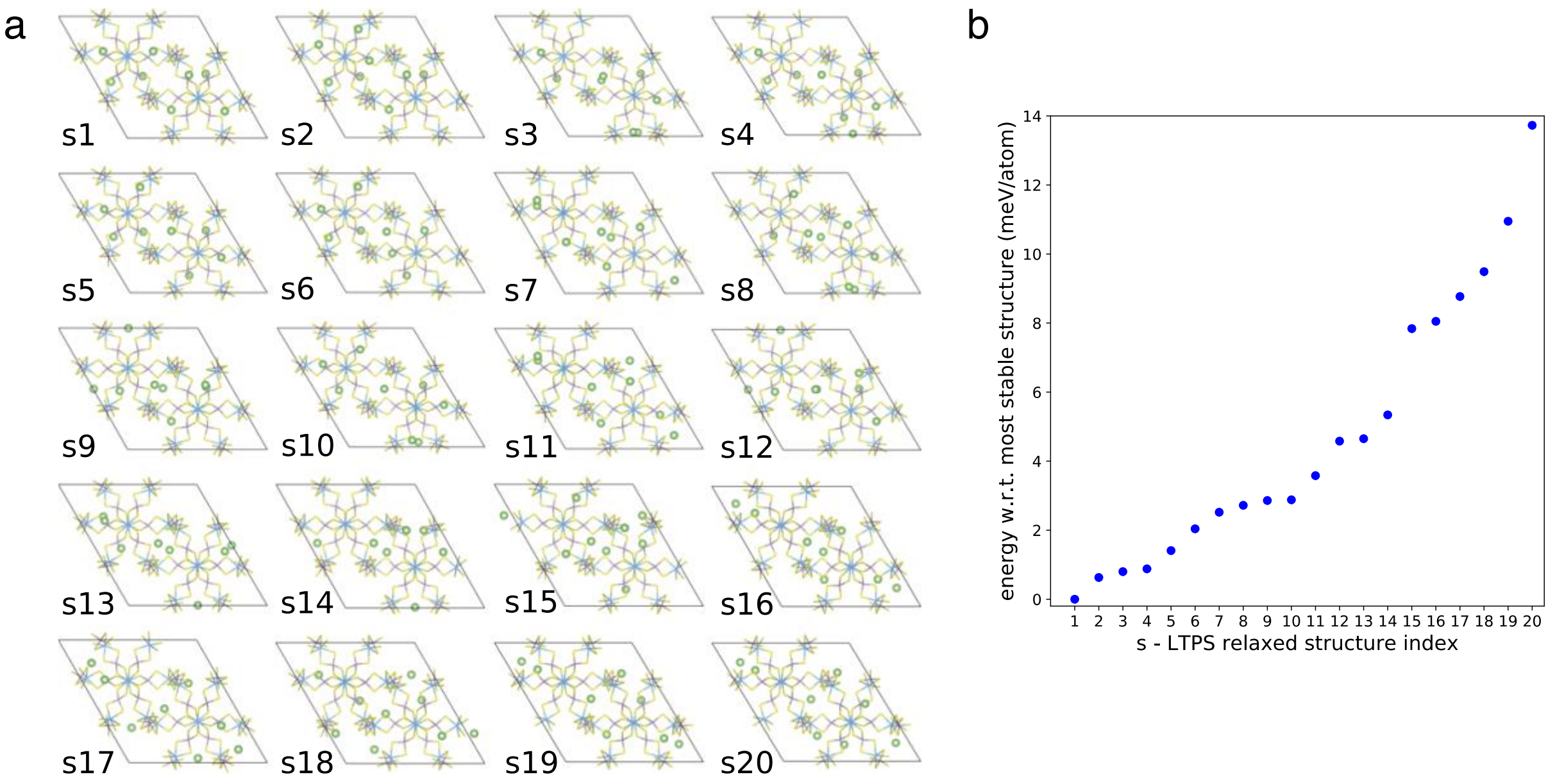

**Fig. S25| LTPS most stable structures computed with DFT** (**a**) Li relaxed positions in the 20 most stable LTPS computed structures and (**b**) their energy with respect to the lowest energy structure.

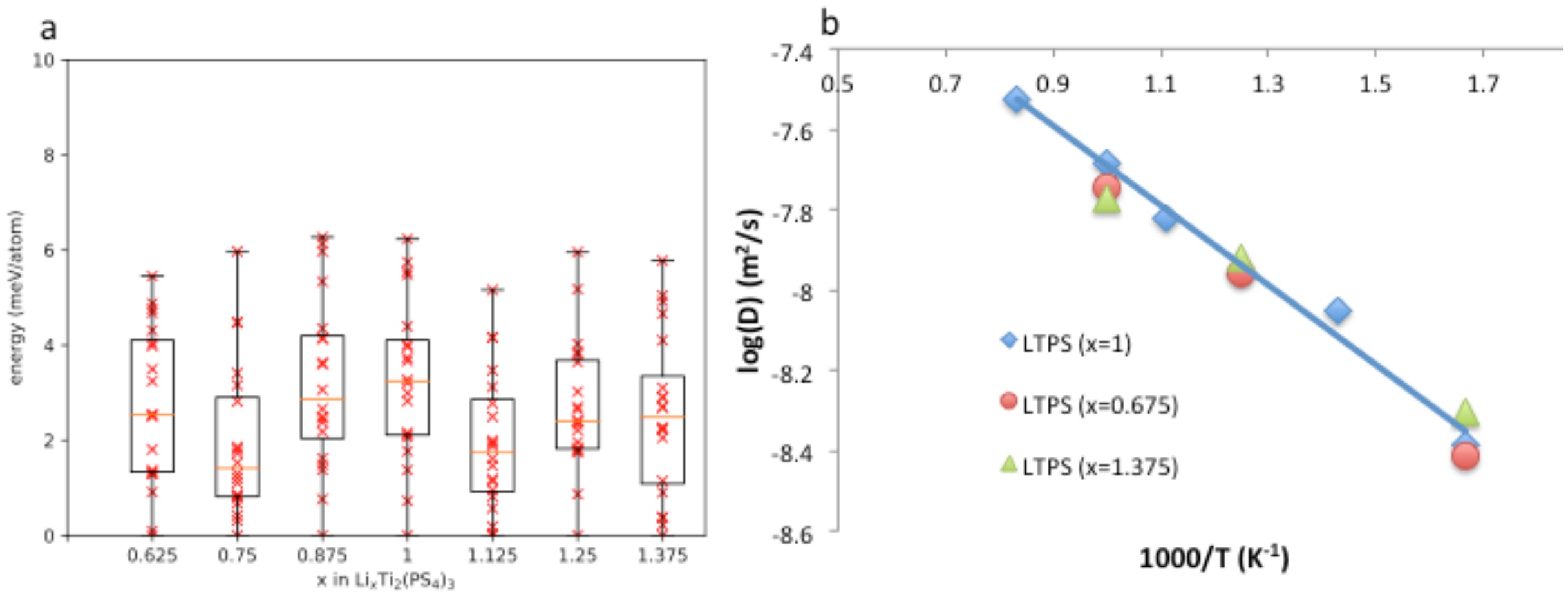

**Fig. S26| Effect of the Li content on the energy landscape and tracer diffusion in LTPS (a)** Distribution of energies obtained by sampling lithium sites with different configurations of lithium. A range of concentration from $Li_{0.675}Ti_2(PS_4)_3$ to $Li_{1.375}Ti_2(PS_4)_3$ was studied. Twenty configurations were chosen for each concentration. The energy spread is small (a few meV/atom) indicating the frustration of the energy landscape. No dependence of the energy distribution on the amount of lithium is observed **(b)** Tracer diffusion computed from AIMD at x=0.625 and x=1.375 (in $Li_xTi_2(PS_4)_3$) obtained after 40ps of computation and compared to the x=1 LTPS computed tracer diffusion. No dependence of the diffusion with the Li content is observed. A charge was added or removed in all computations to keep the formal oxidation states of the other ions. The absence of change in energy landscape and diffusion coefficient with Li content indicates that the Li-Li interactions are not responsible for the energy frustration in LTPS.

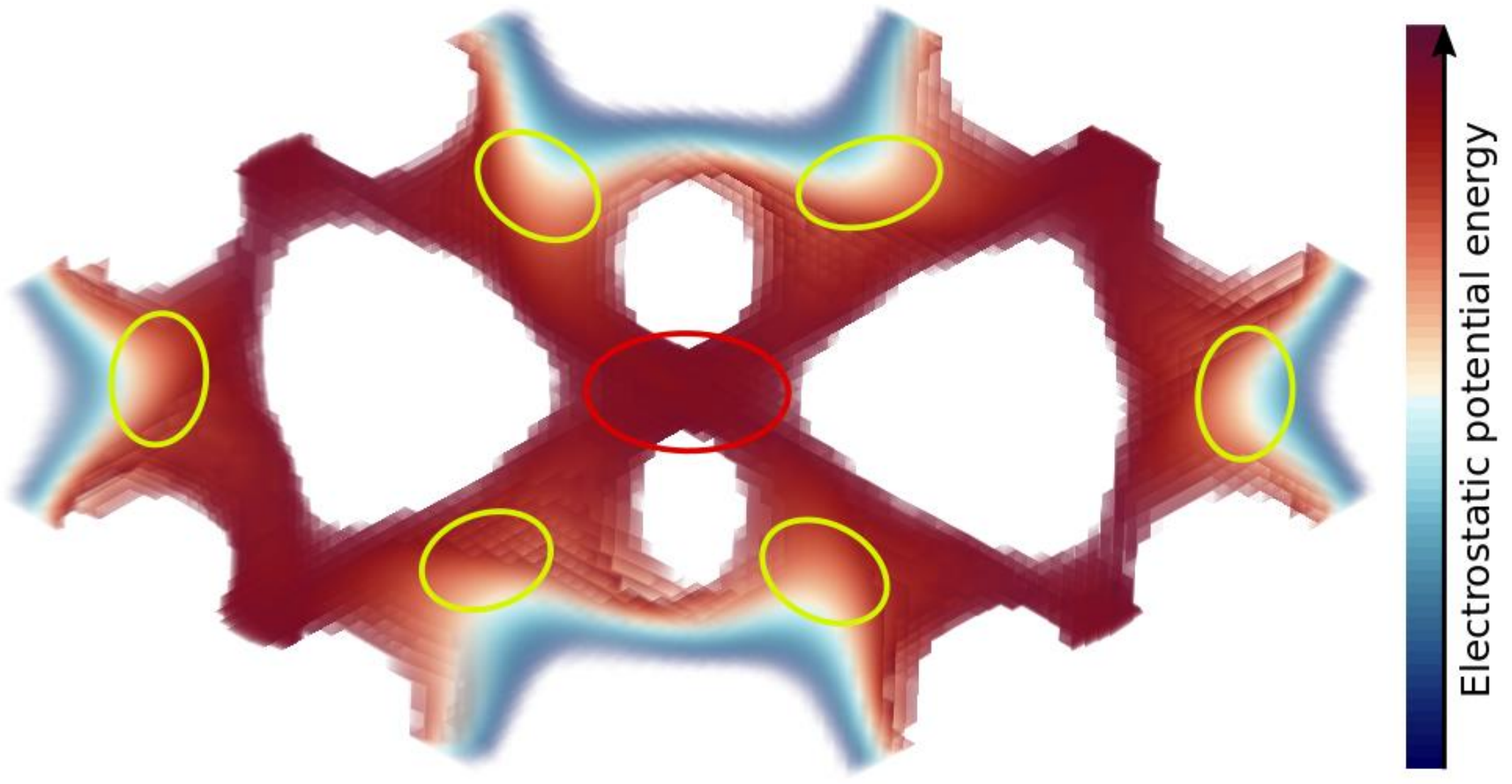

**Fig. S27| Electrostatic potential energy from Li-cation interaction**. The yellow circles indicate the six pockets of high lithium probability of presence (in the AIMD see **Fig. S9**). The red oval indicates the limiting transition state region for inter-ring diffusion.

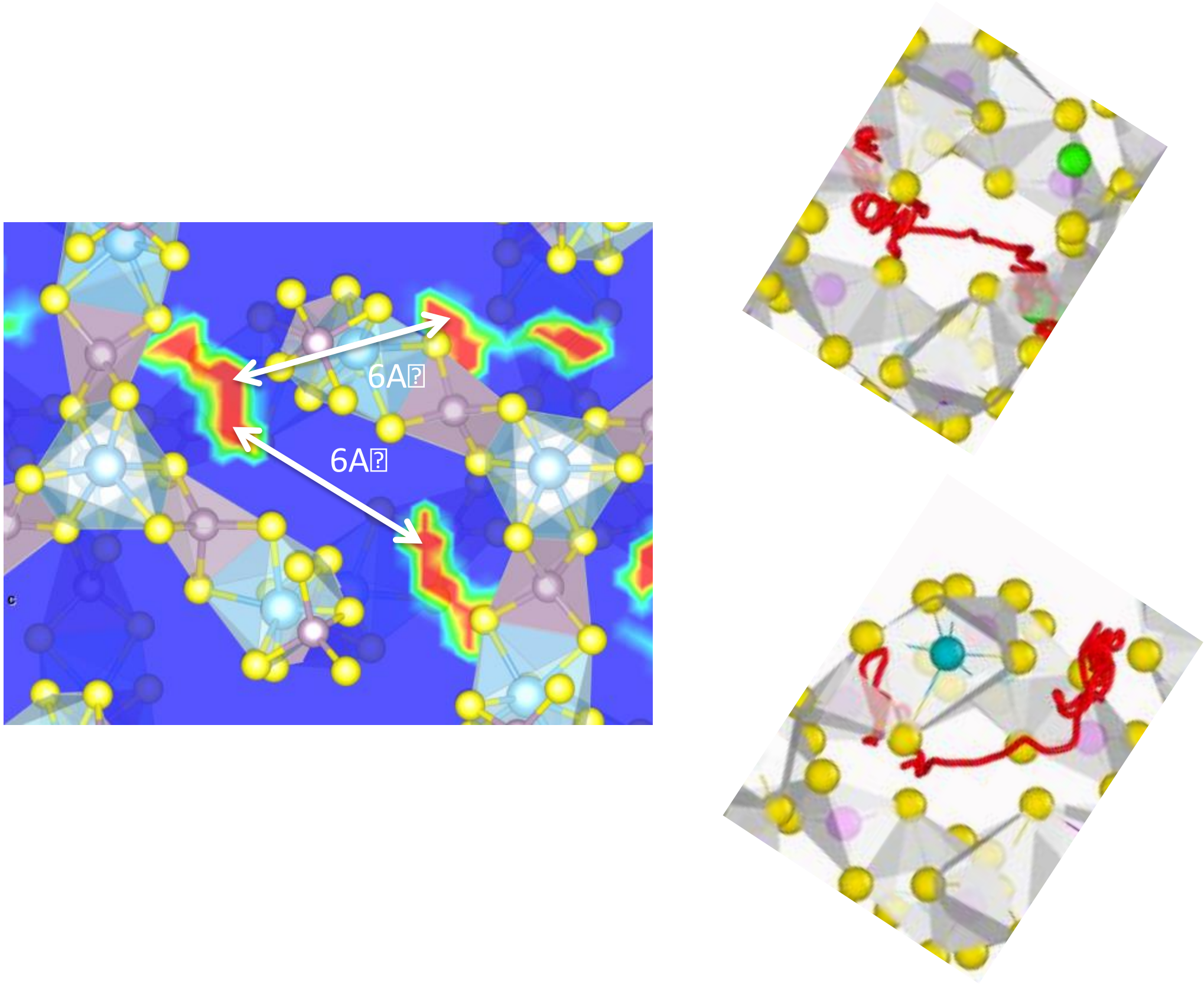

**Fig. S28| Lithium inter-ring jump length and trajectories.** The 6A jump length is estimated from the distance between pockets from 300K AIMD lithium probability density. Two trajectories representing jumps of lithium from one pocket to another are also shown (they come from AIMD at 600K).

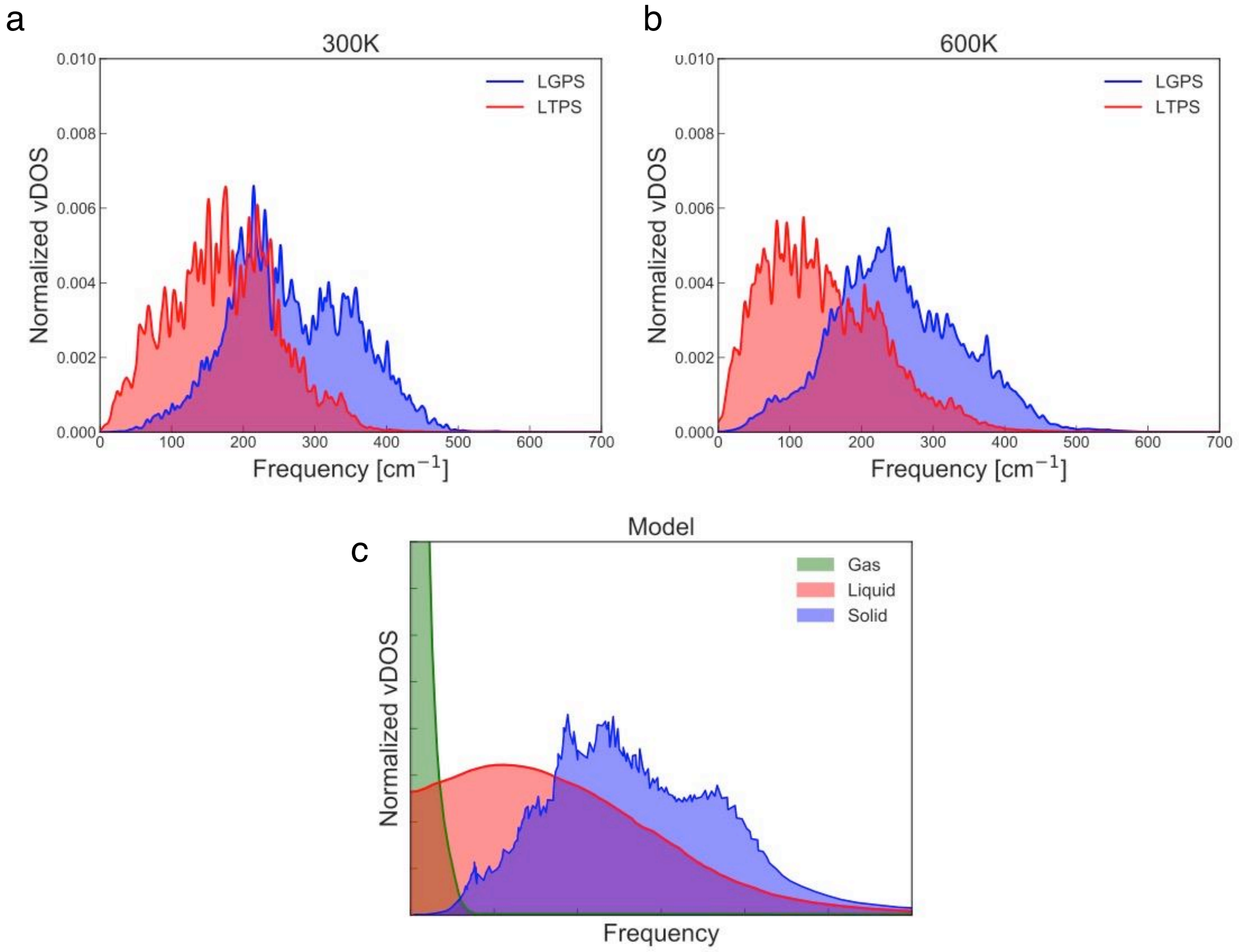

Fig. S29| Vibrational density of states (vDOS) obtained from the lithium velocity auto-correlation function for LTPS, LGPS and model Lennard-Jones systems. (a and b) lithium vDOS at 300K and 600K for LTPS and LGPS. (c) vDOS for a simple Lennard-Jones model with different parameters leading to a solid, liquid or gas behavior [66]. The median vibrational frequency is higher for LGPS than LTPS: respectively 257 cm$^{-1}$ and 173 cm$^{-1}$ at 300K and 241 cm$^{-1}$ and 135 cm$^{-1}$ at 600K. The distribution of the vDOS is also different and can be indicative of the liquid or solid nature of diffusion. Comparing LTPS to LGPS, we observe that LTPS shows a vDOS closer to a liquid than LGPS.

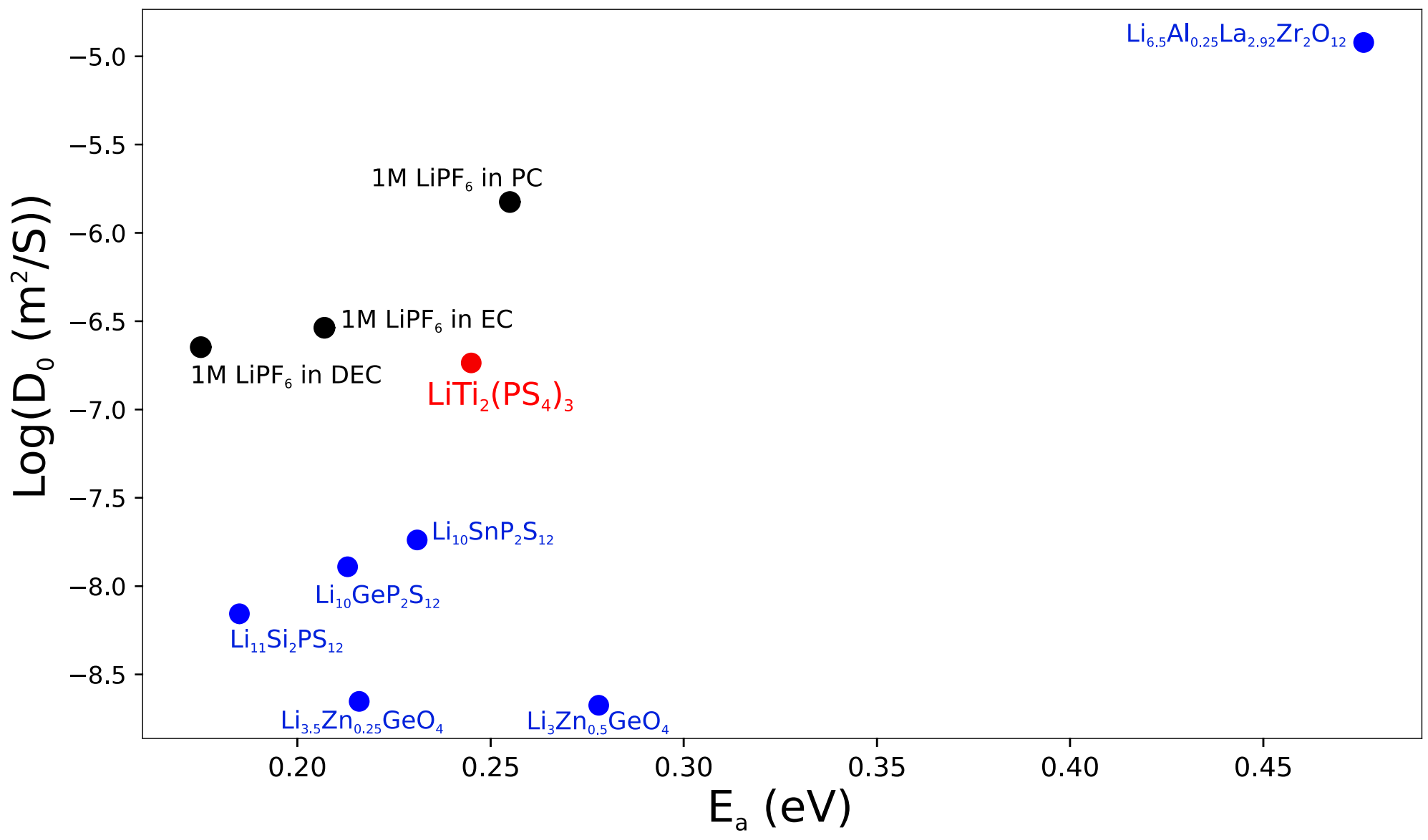

Fig. S30 Pre-factor vs energy barrier of the lithium tracer diffusion coefficient for a series of lithium solid and liquid electrolytes. All data comes from PFG-NMR measurement and from references indicated in Fig. 1. A general dependence between pre-factor and energy barrier is present (Meyer-Neldel rule). The solid lithium superionic conductors offer less attractive combination of pre-factor and energy barriers than liquids. The energy barrier and pre-factor of LTPS sits in between the solids and liquids.

Table S1| Tracer diffusion coefficient of LTPS measured by PFG-NMR

| Temperature, $t$ / ºC | Tracer diffusion coefficient, $D_{Tr}$ / $m^2s^{-1}$ |
|---|---|
| −20 | 0.28 ± 0.012 |
| −10 | 0.34 ± 0.011 |
| 0 | 0.50 ± 0.009 |
| 25 | 1.3 ± 0.058 |
| 50 | 2.9 ± 0.049 |
| 80 | 6.9 ± 0.127 |

**Table S2| Crystal data and structure refinement for $LiTi_2(PS_4)_3$ at 300K**

| | |
|---|---|
| Empirical formula | $LiP_3S_{12}Ti_2$ |
| Formula weight | 1160.74 |
| Temperature | 297(2) K |
| Wavelength | 0.71073 Å |
| Crystal system | Orthorhombic |
| Space group | Ccc2 |
| Unit cell dimensions | a = 19.9565(10) Å |
| | b = 34.6416(19) Å |
| | c = 11.5405(5) Å |
| Volume | 7978.2(7) Å3 |
| Z | 16 |
| Density (calculated) | 1.933 Mg/m3 |
| Absorption coefficient | 2.273 mm-1 |
| F(000) | 4544 |
| Crystal size | 0.200 x 0.060 x 0.060 mm3 |
| Theta range for data collection | 2.944 to 25.225°. |
| Reflections collected | 6962 |
| Independent reflections | 6962 [$R_{int}$ = 0.000]† |
| Completeness to theta = 25.225° | 99.0 % |
| Absorption correction | Semi-empirical from equivalents |
| Max. and min. transmission | 1.000 and 0.975 |
| Data / restraints / parameters | 6962 / 4 / 312 |
| Goodness-of-fit on F2 | 1.086 |
| Final R indices [I>2sigma(I)] | $R_1$ = 0.0424, $wR_2$ = 0.0986 |
| R indices (all data) | $R_1$ = 0.0494, $wR_2$ = 0.1049 |
| Flack parameter | -0.04(3) |
| Largest diff. peak and hole | 0.499 and -0.469 e.Å-3 |

† The equivalents were not merged in order to model the merohedral twinning.

Table S3| Crystal data and structure refinement for $LiTi_2(PS_4)_3$ at 150K

| | |
|---|---|
| Empirical formula | $LiP_3S_{12}Ti_2$ |
| Formula weight | 1160.74 |
| Temperature | 150(2) K |
| Wavelength | 0.7149 Å |
| Crystal system | Orthorhombic |
| Space group | Ccc2 |
| Unit cell dimensions | *a* = 19.7588(8) Å |
| | *b* = 34.0844(13) Å |
| | *c* = 11.3893(2) Å |
| Volume | 7670.3(5) $Å^3$ |
| Z | 16 |
| Density (calculated) | 2.010 $Mg/m^3$ |
| Absorption coefficient | 2.375 $mm^{-1}$ |
| F(000) | 4544 |
| Crystal size | 0.15 x 0.04 x 0.03 $mm^3$ |
| Theta range for data collection | 2.073 to 32.818°. |
| Index ranges | -25<=h<=25, -49<=k<=49, -16<=l<=16 |
| Reflections collected | 12688 |
| Independent reflections | 12688 [$R_{int}$ = 0.000]† |
| Completeness to theta = 25.400° | 99.7 % |
| Absorption correction | Semi-empirical from equivalents |
| Max. and min. transmission | 1.00000 and 0.89846 |
| Data / restraints / parameters | 12688 / 13 / 330 |
| Goodness-of-fit on $F^2$ | 1.094 |
| Final R indices [I>2sigma(I)] | $R_1$ = 0.0504, $wR_2$ = 0.1304 |
| R indices (all data) | $R_1$ = 0.0561, $wR_2$ = 0.1361 |
| Absolute structure parameter | 0.00134(2) |
| Largest diff. peak and hole | 0.804 and -0.942 $e·Å^{-3}$ |

† The equivalents were not merged in order to model the merohedral twinning.

Table S4| Electron and nuclear density modeled inside the channels of $LiTi_2(PS_4)_3$.

| Label | Scattering length at θ = 0 | x | y | z | $U_{eqv}$ Å$^2$ |
|---|---|---|---|---|---|
| single crystal XRD data, not applying the Squeeze procedure | | | | | |
| X1 | 12.8(8) e | 0.5260(12) | 0.0430(6) | 0.539(3) | 0.27(2) |
| X2 | 12.8(8) e | 1/2 | 0 | 0.303(3) | 0.63(5) |
| SR-XRD | | | | | |
| X1 | 68(3) e | 0.572(4) | 0.002(6) | 0.543(13) | 2.38 |
| X2 | 43(4) e | 1/2 | 0 | 0.326(12) | 1.29 |
| NPD | | | | | |
| X1 | 5.7(9) fm | 0.571 | 0.002 | 0.541 | 1.67 |
| X2 | 8.7(4) fm | 1/2 | 0 | 0.308(3) | 0.54 |